\documentclass[10pt, aps, prc, nofootinbib,
amsmath, amssymb, twocolumn,
preprintnumbers, showpacs,
raggedbottom, superscriptaddress,
floatfix]{revtex4-1}

\usepackage[arXiv,todo]{my_paper}  

\providecommand{\exclude}[1]{}
\providecommand{\abs}[1]{\lvert{#1}\rvert}
\providecommand{\norm}[1]{\lVert#1\rVert}
\providecommand{\vect}[1]{\vec{\bm{#1}}}
\providecommand{\op}[1]{\widehat{\bm{#1}}}
\providecommand{\mat}[1]{\bm{#1}}
\renewcommand{\d}{\mathrm{d}}

\providecommand{\braket}[1]{\mathinner{\langle{#1}\rangle}}{\catcode`\|=\active
  \gdef\Braket#1{\left<\mathcode`\|"8000\let|\bravert {#1}\right>}}

\providecommand\accentedsymbol[2]{\def#1{#2}}
\renewcommand\accentedsymbol[2]{\def#1{#2}}

\accentedsymbol{\vq}{\vect{q}}
\accentedsymbol{\vk}{\vect{k}}
\accentedsymbol{\vp}{\vect{p}}
\accentedsymbol{\valpha}{\vect{\alpha}}
\accentedsymbol{\vbeta}{\vect{\beta}}
\accentedsymbol{\vgamma}{\vect{\gamma}}
\accentedsymbol{\oa}{\op{a}}
\accentedsymbol{\ob}{\op{b}}
\accentedsymbol{\oK}{\op{K}}

\usepackage[nameinlink]{cleveref} 

\usepackage{graphicx}
\graphicspath{{./figures/}}

\newcommand\be{\begin{equation}}
\newcommand\ee{\end{equation}}
\newcommand\bea{\begin{eqnarray}}
\newcommand\eea{\end{eqnarray}}

\def\f0{f_{0}}
\def\t0{\tau_{0}}

\newcommand\lsim{\mathrel{\rlap{\lower4pt\hbox{\hskip1pt$\sim$}}
    \raise1pt\hbox{$<$}}}                
\newcommand\gsim{\mathrel{\rlap{\lower4pt\hbox{\hskip1pt$\sim$}}
    \raise1pt\hbox{$>$}}}                

\newcommand{\code}[1]{\verb~#1~}

\usepackage{my_acronyms}
\usepackage{comment}
\graphicspath{{figures/}{figures/arXiv/}}

\newcommand{\E}{\mathcal{E}}    
\newcommand{\xp}{x_p}           

\begin{document}

\title{Constraining the neutron-matter equation of state with gravitational waves}

\newcommand{\UW}{Department of Physics, %
  University of Washington, Seattle, Washington 98195--1560, \mysc{usa}}

\newcommand{\WSU}{%
  Department of Physics \& Astronomy, %
  Washington State University, Pullman, Washington 99164--2814, \mysc{usa}}
\newcommand{\INT}{%
  Institute for Nuclear Theory, University of Washington,
  Seattle, Washington 98195--1560, \mysc{usa}}
\newcommand{\IUCAA}{%
  Inter-University Centre for Astronomy and Astrophysics, %
  Post Bag 4, Ganeshkhind, Pune 411 007, India}
\newcommand{\CMI}{%
  Chennai Mathematical Institute, %
  Siruseri, 603103 India}
\newcommand{\AEI}{%
  Max-Planck-Institut f\"{u}r Gravitationsphysik (Albert Einstein Institute), %
  D-30167 Hannover, Germany}
\newcommand{\LU}{%
  Leibniz Universit\"{a}t Hannover, D-30167 Hannover, Germany}

\author{Michael McNeil Forbes}%
\email{mforbes@alum.mit.edu}%
\affiliation{\WSU}
\affiliation{\UW}

\author{Sukanta Bose}
\email{sukanta@wsu.edu}
\affiliation{\WSU}%
\affiliation{\IUCAA}

\author{Sanjay Reddy}
\email{sareddy@uw.edu}
\affiliation{\INT}

\author{Dake Zhou}
\email{zdk@uw.edu}
\affiliation{\UW}

\author{Arunava Mukherjee}
\email{arunava.mukherjee@aei.mpg.de}
\affiliation{\AEI}%
\affiliation{\LU}

\author{Soumi De}
\email{sde101@syr.edu}
\affiliation{Department of Physics, Syracuse University, Syracuse, New York 13244, \mysc{usa}}

\preprint{\mysc{ligo-p1900097}}
\preprint{\mysc{int-pub-19-009}}

\begin{abstract}
  We show how observations of \gls{GW}s from \gls{BNS} mergers over the next few years can be combined with insights from nuclear physics to obtain useful constraints on the \gls{EoS} of dense matter, in particular, constraining the neutron-matter \gls{EoS} to within 20\% between one and two times the nuclear saturation density $n_0\approx \SI{0.16}{fm^{-3}}$.
  Using Fisher information methods, we combine observational constraints from simulated \gls{BNS} merger events drawn from various population models with independent measurements of the neutron star radii expected from x-ray astronomy (the \gls{NICER} observations in particular) to directly constrain nuclear physics parameters.
  To parameterize the nuclear \gls{EoS}, we use a different approach, expanding from pure nuclear matter rather than from symmetric nuclear matter to make use of recent \gls{QMC} calculations.
  This method eschews the need to invoke the so-called parabolic approximation to extrapolate from symmetric nuclear matter, allowing us to directly constrain the neutron-matter \gls{EoS}.
  Using a principal component analysis, we identify the combination of parameters most tightly constrained by observational data.
  We discuss sensitivity to various effects such as different component masses through population-model sensitivity, phase transitions in the core \gls{EoS}, and large deviations from the central parameter values. 
\end{abstract}

\pacs{}

\maketitle
\tableofcontents

\section{Introduction}
The detection of gravitational waves from the binary neutron star (\gls{BNS}) merger \gls{GW170817} by the \gls{aLIGO} detectors~\cite{TheLIGOScientific:2014jea} in Hanford, \mysc{wa} (\mysc{lho}) and Livingston, \mysc{la} (\mysc{llo}) and the \gls{Virgo} detector~\cite{TheVirgo:2014hva} ushered in the era of multi-messenger astronomy with gravitational waves~\cite{TheLIGOScientific:2017qsa, GBM:2017lvd}. This has been instrumental in launching novel ways of constraining cosmological parameters~\cite{Abbott:2017xzu,Chen:2017rfc,Nair:2018ign,Soares-Santos:2019irc}, on the one hand, and neutron star equation of state (\gls{EoS}) parameters, on the other hand~\cite{TheLIGOScientific:2017qsa, Abbott:2018exr}.
In a \gls{BNS} system the \gls{NS} masses and their \gls{EoS} determine how much quadrupolar deformation $\mathcal{Q}_{ij}$ their tidal fields ${\cal E}_{ij}$ are able to induce in each other.
The two are related by the tidal deformability parameter $\lambda$ as $\mathcal{Q}_{ij} = -\lambda \mathcal{E}_{ij}$.
It is now well understood that the tidal deformability parameters of both \glspl{NS} in a double neutron star system affect the phase of the gravitational wave signal during the late stages of the inspiral~\cite{Flanagan:2007ix}.

Recent articles that followed discovery of \gls{GW170817} have shown that upper bounds on the dimensionless tidal deformability $\Lambda=\lambda c^{10}/(G M)^5$ of the neutron stars obtained from gravitational wave data analysis provide constraints on the \gls{EoS} of dense matter encountered inside neutron stars~\cite{De:2018uhw, Tews:2018iwm, Abbott:2018wiz}.
This is a great opportunity and challenge for several reasons: neutron rich matter, although relevant for many applications, is not easily accessible in experiments, while theoretical approaches require solving the difficult quantum many-body problem and lack a precise characterization of the underlying interactions.
Observational constraints provide an anchor for nuclear theory in this uncertain regime, allowing one to extrapolate low-density and symmetric properties of nuclear matter to significantly improve constraints on neutron-rich matter at higher densities.

In this article we discuss how we can extract more detailed information about the properties of dense neutron-rich matter and neutron stars during the next few years with more \gls{GW} detections and measurements of neutron star radii expected from x-ray astronomy, and highlight the importance of an informed parameterization of the dense matter \gls{EoS}.
We make the reasonable assumption that all neutron stars are described by the same \gls{EoS}.
Further, modern nuclear Hamiltonians based on chiral effective field theory provide a systematic momentum expansion of two- and many-body nuclear forces.
This, combined with advanced computational methods to solve the non-relativistic quantum many-body problem, now allows us to calculate the \gls{EoS} of pure neutron matter up to nucleon number density $n_{c} \approx 2 n_0$, where $n_0=0.16$ nucleons per \si{fm^{3}} is the average nucleon density inside large nuclei (corresponding to a mass density $\rho_0 \simeq \SI{2.7e14}{g/cm^3}$)~\cite{Tews:2018kmu}.
Interestingly, there is a convergence of different ab initio methods
based on realistic microscopic Hamiltonians that account for two and
three neutron forces~\cite{Gandolfi:2009, *Gandolfi:2010b, *Gandolfi:2012, *Gandolfi:2014a}.
These calculations suggest that the functional form of the \gls{EoS} of pure neutron in the density interval $0.5n_0$ to $2n_0$ is well determined.
We use this information to parameterize the \gls{EoS} and show how it helps with the analysis of multiple \gls{BNS} detections and provide tighter and more useful constraints for dense matter physics.
In turn, these constraints for the \gls{EoS} of pure neutron in the density interval where calculations are feasible will provide new insights for nuclear physics.

Our study differs from earlier work in the following aspects:
\begin{itemize}
\item We incorporate insights about neutron-rich matter obtained from nuclear physics by implementing a new parameterization of nuclear equation of state and identify parameters that can be best constrained by \gls{GW} observations.
\item We quantify how constraints on these parameters and on the pressure of neutron matter in the density interval $n_0$ to $2n_0$ will improve with the number of detections.
\item Our analysis uses a numerical relativity based tidal waveform model.
\item We study the effect of different population synthesis models on the accuracy with which \gls{EoS} parameters can be measured with \glspl{GW} and use several thousand binary neutron star source simulations to assess errors in \gls{EoS} parameter measurements.
\item While a nearby event like \gls{GW170817} at \gls{aLIGO} design sensitivity would significantly constrain the properties of neutron matter, we show that similar constraints can be obtained from about 15 events beyond \SI{100}{Mpc}.
\end{itemize}

We begin with a summary of our results in \cref{sec:results}, then describe how we have parameterized the dense matter \gls{EoS} in \cref{sec:eos}.
In \cref{sec:GW} we discuss how we obtain constraints from the gravitational waveform of simulated merger events.
Finally, we discuss details of the method we use to obtain these constrains in \cref{sec:stat}.

\begin{figure*}[t]
  \includegraphics[width=\textwidth]{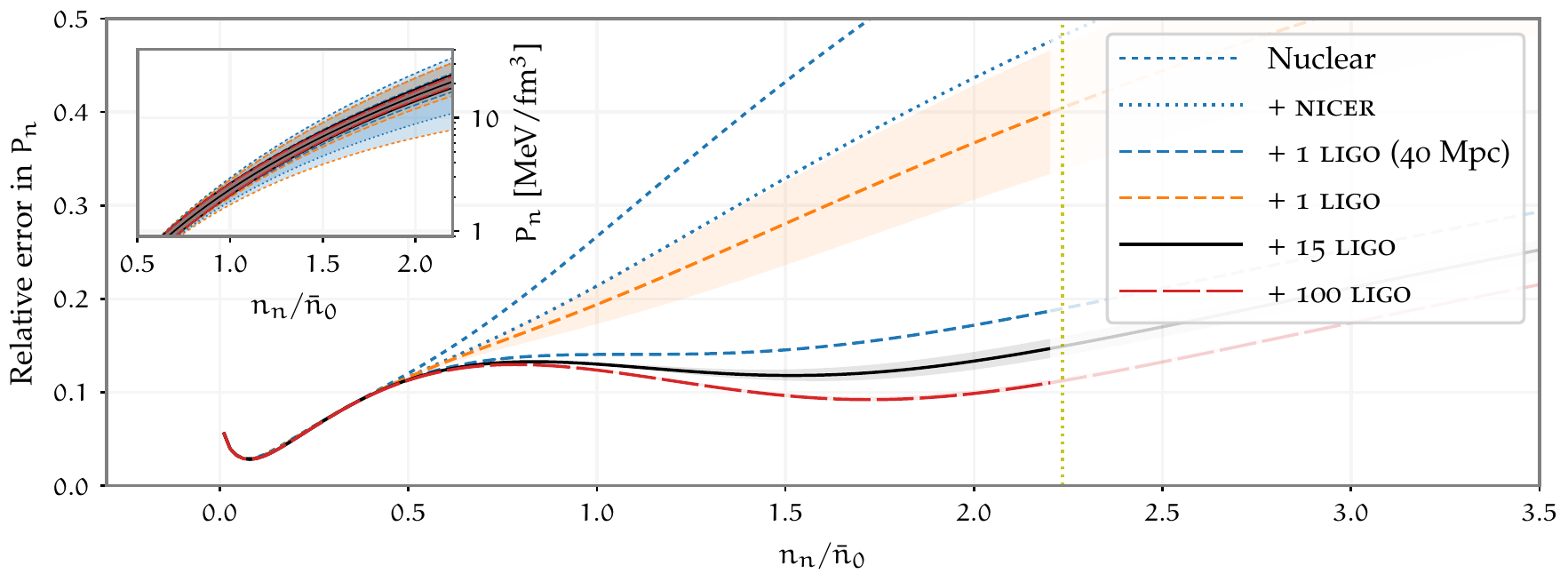}
  \caption{\label{fig:progressive_constraints}\glsreset{EoS}%
    (color online)
    Relative constraints on the pressure of neutron matter $P_n(n_n)$ from $N_{\mathrm{obs}} \in \{1, 15, 100\}$ simulated merger events, and expected constraints from \gls{NICER} (J0437)~\cite{Miller:2016a} ($M=\num{1.44+-0.07}M_\odot$, $\delta R/R = \num{0.1}$).
    From top: constraints from nuclear theory augmented by \gls{NICER}, from a single merger event at $D=\SI{40}{Mpc}$ with \gls{aLIGO} sensitivity, then various $N_{\mathrm{obs}}$ \gls{LIGO} events drawn from \gls{CompleteStdAsubsolarNSNS} that have $D \in [\SI{100},~\SI{400}]{Mpc}$.
    The shading shows the range of sampling errors ($1\sigma$ or 68th percentile) demonstrating variation within the \gls{CompleteStdAsubsolarNSNS} population model~\cite{Dominik:2012kk}.
    Beyond the vertical yellow line, we use the core \gls{EoS}.
    Inset: $P_n(n_n)$ with $1\sigma$ error bands corresponding to each of the constraints.}
\end{figure*}

\section{Results}
\label{sec:results}
Our main result is that even a handful of \gls{GW} observations of \gls{BNS} mergers will provide the most stringent constraints on the low-temperature equation of state of dense neutron matter in the density interval between $n_0-2 n_0$.
This is summarized in \cref{fig:progressive_constraints}, which shows how the constraints on the pressure of pure neutron matter $P_n(n_n)$ improve as a function of additional \gls{NICER} or \gls{LIGO} observations.
We start from the errors listed in \cref{tab:Central2c}, which, for the purposes of this analysis, we interpret as uncorrelated $1\sigma$ normal errors for the parameters.
This gives the upper dotted line labeled ``Nuclear''.

To this, we add the following constraints:
\begin{itemize}
\item Constraints from a simulated binary with similar masses and distance $D\sim\SI{40}{Mpc}$ to \gls{GW170817} but at \gls{aLIGO} design sensitivity. 
\item \Gls{GW} observations at \gls{aLIGO} design sensitivity of $N_{\mathrm{obs}} \in \{1, 15, 100\}$ distant $D \in [\SI{100},~\SI{400}]{Mpc}$ simulated merger events from population model \gls{CompleteStdAsubsolarNSNS} as described in \cref{sec:GW}.
  To estimate the variance possible within the population model, we sample 500 different populations, each containing $N_{\mathrm{obs}}$, and plot the $1\sigma$ (68th percentile) error bands as shaded regions.
\item An uncorrelated mass and radius measurement of \mysc{j0437} projected to be measured at a 5\% level from \gls{NASA}'s \gls{NICER} mission --
  i.e. $\num{1.44+-0.07}M_\odot$ with a 10\% measurement of $R$~\cite{Miller:2016, Miller:2016a}.
\end{itemize}

\begin{figure}[tb]
  \includegraphics[width=\columnwidth]{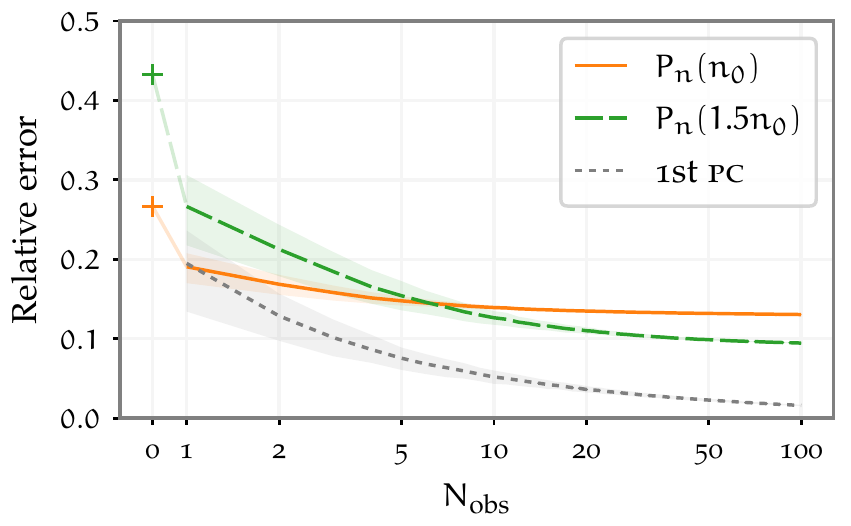}
  \caption{\label{fig:pc}%
    (color online) Improvement in relative constraint on the pressure of neutron matter $P_n(n_n)$ at $n_n=1.5n_0$ and $n_n=n_0$ (related to the slope $L = 3P_n(n_0)/n_0$ of the symmetry energy) to an increasing number of simulated merger events applied to the initial nuclear constraints denoted with a plus at $N_{\text{obs}} = 0$.
    The shading shows the range of sampling errors ($1\sigma$ or 68th percentile) demonstrating variation within the \gls{CompleteStdAsubsolarNSNS} population model.
    The lower dotted curve shows the level of the most tightly constrained principal component (1st \mysc{pc}).
  }
\end{figure}

This analysis demonstrates several key points:
A nearby event such as \gls{GW170817} is comparable to a dozen or so events from $D\geq \SI{100}{Mpc}$.
The \gls{NICER} constraints are comparable to a single \gls{LIGO} observation from a distant population sample having low \gls{SNR}, however, nearby or multiple accumulated \gls{LIGO} events yield significant improvement.
After about $N_{\mathrm{obs}}=15$ observation, we observe rather limited improvement from additional $N_{\mathrm{obs}}=100$.
This can also be seen in \cref{fig:pc}, which shows how the constraints improve as a function of the number of observations.

One caveat: these constraints assume Gaussian errors and linear error propagation.
A proper analysis requires a much more expensive Bayesian approach (see, e.g., Ref.~\cite{Agathos:2015uaa}).
To assess the non-linear effects, we provide similar plots for comparison in~\cite{EPAPS} for the different central values listed in \cref{tab:eos}.

To put these results in perspective, consider the nuclear symmetry energy $S_{\mathrm{sym}}$ and the slope of its density dependence $L$,
\begin{subequations}
  \begin{align}
    S_{\mathrm{sym}} & = E_{np}(n_0, 0) - E_{np}(\tfrac{n_0}{2}, \tfrac{n_0}{2}), \label{eq:S}\\
    L                & = 3n_0\left.\pdiff{E_{np}(n_n, 0)}{n_n}\right|_{n_n=n_0} = 3\frac{P_n(n_0)}{n_0},\label{eq:L}
  \end{align}
\end{subequations}
where $E_{np}(n_n, n_p)$ is the energy-per-particle of uniform nuclear matter.
If the so-called parabolic approximation holds at saturation ($L_2 \approx L$ -- see \cref{eq:E_np} and the surrounding discussion), then upcoming neutron skin experiments~\cite{Horowitz:2012, *Horowitz:2014, *Horowitz:2014a} expect to constrain $\Delta L = \SI{41}{MeV}$ with a possible reduction to $\Delta L = \SI{15}{MeV}$ with a followup experiment.
This is comparable to combined constraints from ab initio calculations~\cite{Hebeler:2010, *Hebeler:2013, Wlazlowski:2014a, Gandolfi:2014, Lynn:2015} and astrophysical observations~\cite{Page:2006, Gandolfi:2009, *Gandolfi:2010b, *Gandolfi:2012, *Gandolfi:2014a, Steiner:2012, *Steiner:2013, Lattimer:2014}.
From our analysis we thus see that \gls{GW} observations alone could have an impact at the $\sim 15\%$ level corresponding to $\Delta L \approx \SI{10}{MeV}$.

\begin{figure}[htbp]
  \glsreset{EoS}
  \includegraphics[width=\columnwidth]{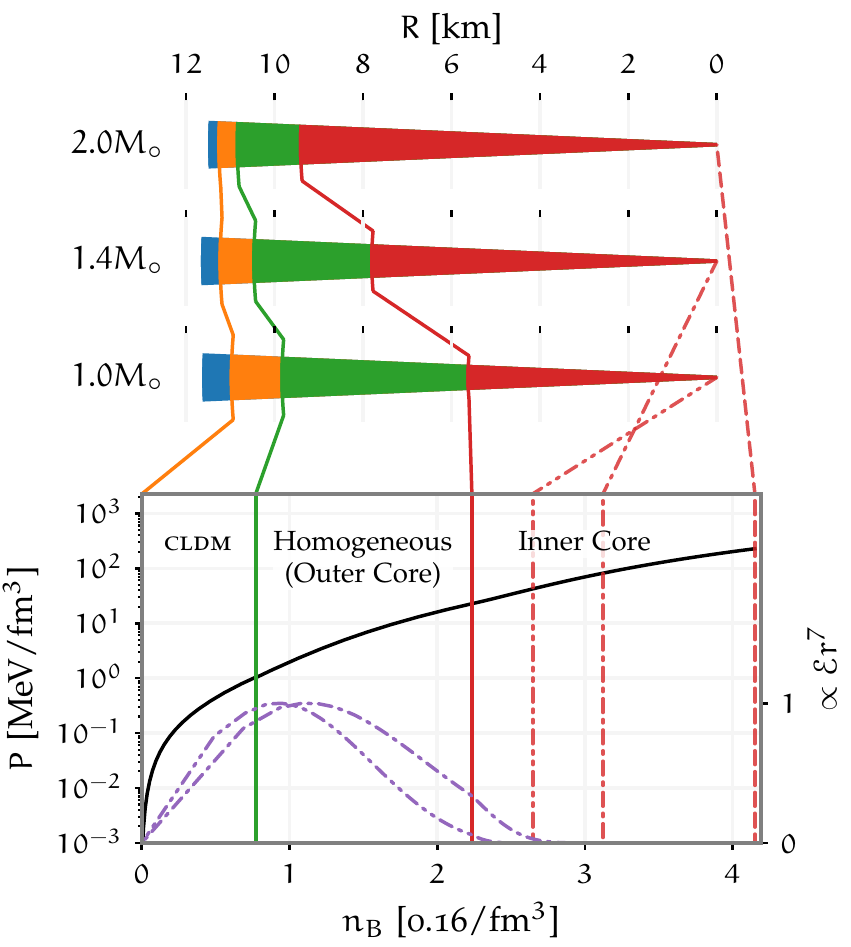}
  \caption{\label{fig:eos}%
    (color online) Regions of the neutron star.
    The upper three wedges represent a cross-section of $M=2M_\odot$, $M=1.4M_\odot$, and $M=1M_\odot$ neutron stars respectively.
    As discussed in the text, the unified \acrshort{EoS} smoothly connects four distinct regions from low density on the left to high density on the right.
    The radius of these transitions for the \gls{Central2c} parameter values is shown in the top plot.
    These are connected to the equation of state expressed in terms of the pressure $P(n_B)$ (solid (black) line on left axis) as a function of the total baryon density in units of the saturation density $n_0=\SI{0.16}{fm^{-3}}$.
    From low to high density, the regions of the \acrshort{EoS} are: a) the outer crust (very low density which is too small to see on the lower plot) that interpolates the data of~\cite{Baym:1971} and~\cite{Negele:1973} as tabulated in~\cite{Sharma:2015} (blue) with minor corrections to ensure convexity as discussed in~\cite{EPAPS}; b) the inner crust modeled by the \acrshort{CLDM}~\cite{Haensel:2007, *Chamel:2008} (orange); c) the outer core of homogeneous nuclear matter in beta-equilibrium (green); d) the inner core equation of state parameterized by a quadratic speed of sound (red).
    At the right, the various (red) dashed lines correspond to the core density of the respective stars.
    At the bottom are corresponding dashed curves (purple) proportional to $\mathcal{E}r^7$ (normalized to the maximum value on the right axis) for the two lower-mass stars.
    This roughly correlates with the local contribution to the dimensionless tidal deformability~\cite{Nelson:2018xtr}.}
\end{figure}

\section{Parameterization of the Nuclear Equation of State}
\label{sec:eos}
\glsreset{EoS}
To relate the nuclear equation of state to the structure of neutron stars, we must first characterized the \gls{EoS} of nuclear matter.
This is conveniently parameterized by the energy density $\mathcal{E}(n_B)$ as a function of the baryon number density $n_B = n_n + n_p$, which is the sum of the neutron and proton number densities.
Simple approximation for this function in terms of polytropes are often a starting point for astrophysical analysis.
Indeed, many families of nuclear \gls{EoS} can be characterized quite well by a simple set of piecewise polytropes~\cite{Read:2009}.

Our approach here, however, is to directly express $\mathcal{E}(n_B)$ in terms of nuclear physics parameters. This approach allows one to directly assess how observations translate into constraints on nuclear physics.
We shall demonstrate this by providing constraints on the pressure of pure neutron matter $P_n(n_n)$, which is inaccessible from a general polytropic analysis (\cref{fig:progressive_constraints}).

It is useful to divide the neutron star interior into four regions: the outer crust, the inner crust, the outer core, and the inner core.
The radial extent of the outer crust, which is composed neutron-rich nuclei embedded in a electron gas, is only a few hundred meters and its contribution to the neutron star mass is negligible.
The \gls{EoS} of the outer crust is well understood and depends weakly on the composition of nuclei present.
The inner crust extends from $n = n_{\text{drip}}\simeq \num{2e-3}n_0$ to $n=n_{\text{core}} \simeq 2n_0$, has radial thickness $\sim SI{2}{km}$, and contains a modest fraction of the mass.
Here, exotic neutron-rich nuclei are embedded in a dense liquid of neutrons and electrons, as described by the \gls{CLDM} in \cref{sec:cldm}.
The outer core is a liquid composed primarily of neutrons and a small (few percent) admixture of protons, electrons, and muons.
It extends from $n\sim 0.5n_0$ to $n=n_c \sim 2n_0$ where the description of matter in terms of nucleons interacting with static potentials is expected to break down.
The inner core extends to higher densities, and we switch here to the speed-of-sound parameterization discussed in \cref{sec:cs}.

\begin{figure}[t]
  \includegraphics[width=\columnwidth]{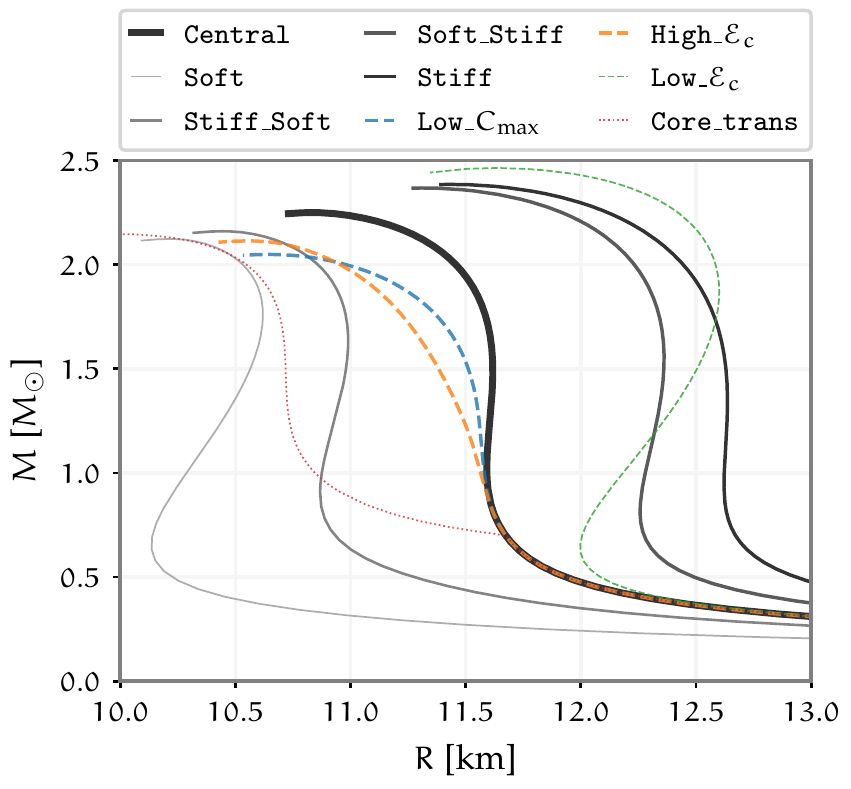}
  \caption{\label{fig:MR}%
    (color online) Mass-radius curves for the \glspl{EoS} considered in \cref{tab:eos}.
    The thick solid curve is our \gls{Central2c} \gls{EoS}.
    Dashed curves correspond to different core parameterizations.
    Thin curves correspond to \glspl{EoS} for which astrophysical observations would provide poor constraints for nuclear physics.
    These include a sharp first-order transition in the core (\gls{Central2c_trans}), and soft \glspl{EoS} (\gls{Soft2c} and \gls{Stiff2c_Soft}) which form very compact objects with low deformability.
    The \gls{Central2c_low_Ec} \gls{EoS} also poorly constrains nuclear physics since the core appears close to the saturation density.
    As shown later in \cref{fig:PCA_all}, for these types of \gls{EoS}, observations constrain the core parameters rather than the properties of neutron matter.
  }
\end{figure}

On dimensional grounds one expects the dimensionless tidal deformability $\Lambda$ to be related to $\int^R_0 \mathcal{E} r^n\d{r} = \braket{\mathcal{E} r^n}$ with $n\sim 7$ for $M \sim 1.4M_\odot$~\cite{Nelson:2018xtr}.
Although the \gls{EoS} around intermediate densities dominates the 7th moment of energy distribution for massive neutron stars, the inner crust also makes a large contribution to $\Lambda$ for low-mass stars (which are believed to be more common in binary neutron star systems).
This contribution is shown by the dashed (purple) lines at the bottom of \cref{fig:eos}.
Thus, it is important to provide a unified description of the \gls{EoS} of the inner crust and the outer core in any analysis that aims to constrain the \gls{EoS} using \gls{GW} observations of binary neutron stars.

\subsection{Compressible Liquid Drop Model}\label{sec:cldm}
\glsreset{CLDM}
The \gls{CLDM} (see~\cite{Haensel:2007, *Chamel:2008}) provides a unified \gls{EoS} connecting a fixed outer crust for $\rho < \rho_{\text{drip}}$
(for which we use the data in Table 4 of~\cite{Sharma:2015}) to the inner core \gls{EoS}.
In the inner crust, the \gls{CLDM} constructs spherical nuclei in a spherical Wigner-Seitz cell, ensuring equilibrium with surrounding neutron and lepton gases by establishing both electric and $\beta$-equilibrium.
This is similar to the approach taken in~\cite{Fortin:2016, Zdunik:2016}, but differs in how we define the nuclear matter \gls{EoS} $\E_{np}(n_n, n_p)$.
Instead of using $\E_{np}(n_n, n_p)$ obtained from specific models based on effective Hamiltonians solved in the mean field approximation to reproduce empirical parameters like nuclear saturation properties, we use what we believe is close to a minimal phenomenological parameterization that directly encodes properties that can either be measured or calculated reliably.
The advantage of our approach is that these parameters are directly connected with the unified \gls{EoS}, allowing us to provide a full covariance analysis linking nuclear parameters with neutron star observables.

Although the use of a spherical Wigner-Seitz cell precludes the possibility of pasta phases~\cite{Ravenhall:1983uh}
the errors incurred by the Wigner-Seitz approximation for different lattice structures are less then 0.5\% (see e.g.~\cite{Chamel:2007, Chamel:2008}).

Our implementation of the \gls{CLDM} introduces two effective parameters:
the surface tension $\sigma_0$ and the parameter $C_{\text{sym}} = \sigma_\delta/\sigma_0$ which characterizes the isospin dependence of the surface tension $\sigma(n_n, n_p) = \sigma_0 \bigl(1 -C_{\text{sym}}(\beta_p)^2 + \order(\beta_p^4)\bigr)$~\cite{Lattimer:1985} (see~\cite{EPAPS} for the exact form used), where $\beta_p = (n_n-n_p)/(n_n+n_p)$ is the isospin asymmetry.
We fix the parameter $\sigma_0$ to smoothly match the tabulated outer crust equation of state, leaving free the single parameter $\sigma_\delta$.
Additionally, we include as a parameter a suppression factor $\mathcal{C}$ for the Coulomb interaction to allow for the diffusivity of the proton charge distribution (see the discussion in~\cite{Steiner:2012a}). As will be shown in \cref{sec:results}, these parameters have negligible effects on the constructed \gls{EOS}.

This approach allows for a small first-order phase transition from the region modeled by the \gls{CLDM} to homogeneous nuclear matter.
With our parameters, this phase transition is weak: $\delta n < \SI{0.002}{fm^{-3}}$.

To establish $\beta$-equilibrium we include leptons modeled as a Fermi gas of electrons (and muons at sufficiently high densities) in the \gls{TF} approximation.

\begin{table}[tb]
  \begin{tabular}{p{\columnwidth}}
    \toprule
    \Glsentryname{CLDM} parameters:
    {\begin{align*}
      \sigma_\delta & = \SI{1.38+-1.38}{MeV/fm^2},
                    & \mathcal{C}                  & = \num{0.9(1)},     \\
      \intertext{Symmetric nuclear matter and symmetry parameters:}
      n_0           & = \SI{0.16+-0.01}{fm^{-3}},
                    & S_2                          & = \SI{31+-4}{MeV},  \\
      e_0           & = \SI{-16+-0.3}{MeV},
                    & L_2                          & = \SI{60+-40}{MeV}, \\
      K_0           & = \SI{240+-40}{MeV},
                    & K_2                          & = \SI{30+-30}{MeV}, \\
      \intertext{Neutron matter parameters:}
      a             & = \SI{13+-0.3}{MeV},
                    & \alpha                       & = \num{0.5+-0.02},  \\
      b             & = \SI{3.5+-1.5}{MeV},
                    & \beta                        & = \num{2.3+-0.5},
    \end{align*}
    Proton polaron parameters:
    \begin{align*}
      \mu_p(n_0) & = \SI{-105+-10}{MeV},
                 & u_p                        & = \num{3.1+-0.6},
                 & \frac{m_{\text{eff}}}{m_p} & = \num{0.8+-0.1},
    \end{align*}
    Inner-core parameters:
    \begin{align*}
      \mathcal{E}_c & = \SI{350+-35}{MeV},
                    & \mathcal{E}_{\mathrm{max}} & = \SI{0.8+-0.4}{GeV},
                    & C_{\mathrm{max}}           & = \num{0.8+-0.2}.
    \end{align*}
    \vspace{-1.5\baselineskip}} \\
    \bottomrule
  \end{tabular}
  \caption{\label{tab:Central2c} Parameters defining the \gls{Central2c} \gls{EoS} along with their uncorrelated $1\sigma$ covariance (expressed using the \gls{SI} convention $\num{3.5+-1.5}\equiv\num[separate-uncertainty]{3.5\pm 1.5}$) used to defined the ``Nuclear'' error estimates prior to including information from astrophysical observations.
    We take the values of the \gls{CLDM} parameters from the fits to the \mysc{apr} \gls{EoS} tabulated in~\cite{Steiner:2012a} but assign large errors to encompass missing physics such as the possibility of pasta phases.
    Symmetric nuclear matter and symmetry parameters have errors taken from the extensive analysis~\cite{Margueron:2018}.
    Neutron matter parameters have errors estimated from \gls{QMC} calculations with various three-body interactions~\cite{Gandolfi:2009, *Gandolfi:2010b, *Gandolfi:2012, *Gandolfi:2014a}, and are consistent with recent \gls{QMC} results based on chiral \gls{EFT} interactions~\cite{Wlazlowski:2014a, Gandolfi:2014, Gandolfi:2015, Lynn:2015}.
    Proton polaron parameters have errors estimated from the \gls{QMC} calculations~\cite{Roggero:2014} and are consistent with estimates from chiral interactions~\cite{Rrapaj:2016}.
    The core parameters are chosen to allow for a $2M_\odot$ star at the extremes of all of our models except for the \gls{Soft2c} \gls{EoS} which requires a lower core transition and are given large errors to be conservative with the exception of the parameter $\mathcal{E}_c$.
    This is given a small error for the purposes of our statistical analysis as the dependence is highly non-linear.
    Variations of this parameter are considered specially in \cref{fig:core_constraints}.}
\end{table}

\begin{table}[tb]
  \sisetup{
    table-number-alignment = center,
    table-format = 3.1,
    table-unit-alignment = left,
  } %
  
  \robustify\bfseries
  \robustify\itshape
  \begin{tabular}{
    l
    S[table-format = 2.1]
    S[table-format = 1.1]
    S[table-format = 1.1]
    S[table-format = 1.1]
    S[table-format = 3.0]
    S[table-format = 1.1]}
    & \multicolumn{4}{c}{Neutron Matter}
    & \multicolumn{2}{c}{Inner Core}\\
    \toprule
    \gls{EoS}
    & {$a$ [\si{MeV}]} & {$\alpha$} & {$b$ [\si{MeV}]} & {$\beta$}
    & {$\mathcal{E}_c$ [\si{MeV/fm^3}]} & {$C_{\mathrm{max}}$}\\
    \midrule
    \gls{Central2c} & 13.0 & 0.5 & 3.5 & 2.3 & 350 & 0.8 \\
    \midrule
    \gls{Soft2c} & 12.7 & 0.3 & 2 & 2.1 &\\
    \gls{Stiff2c} & 13.3 & 0.7 & 5 & 2.5 &\\
    \gls{Soft2c_Stiff} & 12.7 & 0.3 & 5 & 2.5 &\\
    \gls{Stiff2c_Soft} & 13.3 & 0.7 & 2 & 2.1 &\\
    \gls{Central2c_low_Ec} &&&&& 200 &\\
    \gls{Central2c_high_Ec} &&&&& 500 &\\
    \gls{Central2c_low_C_max} &&&&&& 0.6\\
    \midrule
    &&&&& \multicolumn{2}{l}{$\mathcal{E}_{\mathrm{trans}}$ [\si{MeV/fm^{3}}]}\\
    \gls{Central2c_trans} &&&&&150&\\
    \bottomrule
  \end{tabular}
  \caption{\label{tab:eos}%
    List of changed \gls{EoS} parameters compared in this work.
    All other parameters share the same values as the \gls{Central2c} \gls{EoS} in the top row, which takes the central values listed in \cref{tab:Central2c}.
    The first four variations -- \gls{Soft2c}, \gls{Stiff2c}, \gls{Soft2c_Stiff}, and \gls{Stiff2c_Soft} -- refer to the properties of the neutron-matter equation of state and whether the \gls{EoS} of the outer core is softer or stiffer than \gls{Central2c} at low/high density.
    The next three variations -- \gls{Central2c_low_Ec}, \gls{Central2c_high_Ec}, and \gls{Central2c_low_C_max} -- explore variations of the core \gls{EoS}.
    To better understand the sensitivity of our results to the properties of the core, we include one slightly different form \gls{Central2c_trans} which has a first-order phase transition with discontinuity $\mathcal{E}_{\mathrm{trans}}$.
    (See \cref{fig:core_constraints}.)
  }
\end{table}

\begin{figure*}[t]
  \includegraphics[width=\textwidth]{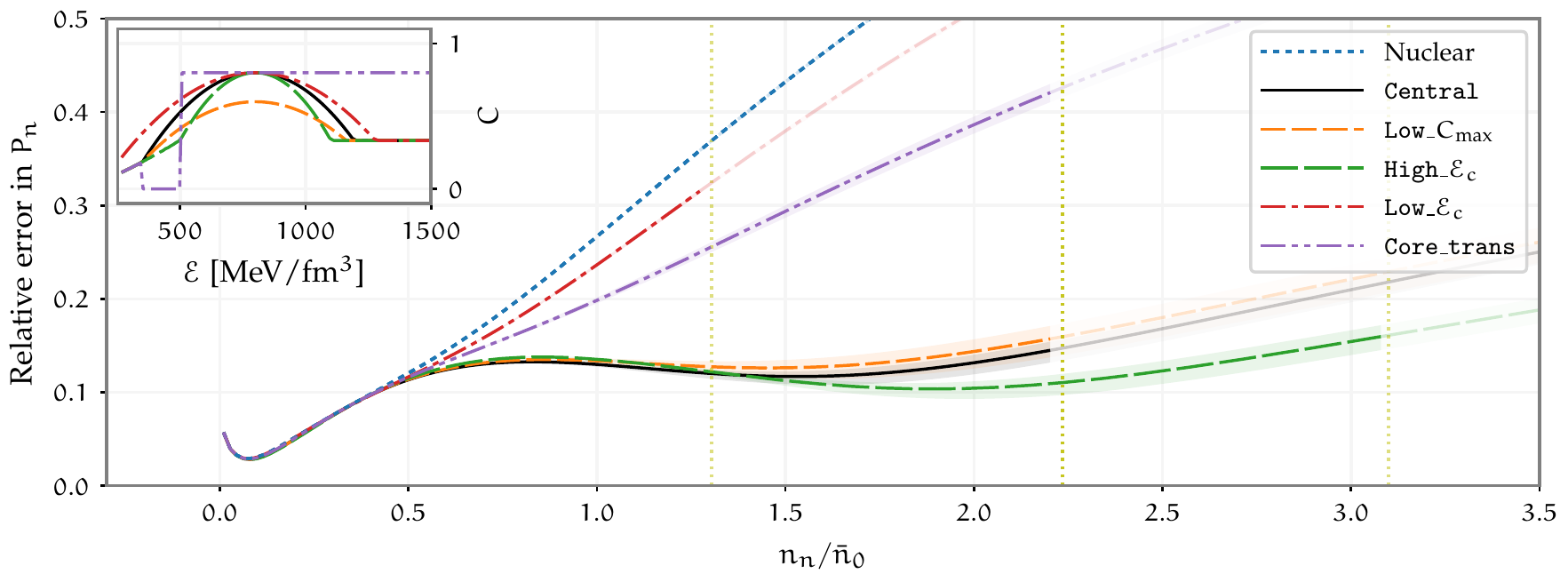}
  \caption{\label{fig:core_constraints}%
    (color online) Sensitivity of the constraint on the pressure of neutron matter $P_n(n_n)$ from $N_{\mathrm{obs}} = 15$ simulated merger events drawn from the \gls{CompleteStdAsubsolarNSNS} population model to variations of the core equation of state.
    The vertical yellow lines denote the density $n_c$ at which the \gls{EoS} reverts to the core form.
    Inset: form of the various core speed-of-sound functions $C(\mathcal{E}) = c_s^2/c^2$.
  }
\end{figure*}

\subsection{Homogeneous Nuclear Matter}
One of the main new features of our analysis is to parameterize the nuclear-matter \gls{EoS} as an expansion in the proton fraction $x_p = n_p/(n_n+n_p)$ from pure neutron matter to symmetric neutron matter.
This is in contrast to the common approach of expanding about symmetric nuclear matter in powers of the isospin asymmetry $\beta_p = (n_n-n_p)/(n_n+n_p)$.
The common approach allows one to directly connect experimentally relevant properties of symmetric nuclear matter to properties of neutron matter.
This connection, however, is generally predicated on the so-called parabolic approximation, which is valid only if quadratic terms $\beta_p^2$ dominate over quartic $\beta_p^4$ and higher-order terms.
While there is some support for this below saturation density from relativistic \gls{DBHF} calculations~\cite{Lee:1998}, Gogny forces~\cite{Gonzalez-Boquera:2017}, and other perturbative techniques (see~\cite{Li:2008} for a review), it is not well established at higher densities.
Indeed virtually any form of neutron-matter \gls{EoS} can be accommodated with quartic $\beta_p^4$ terms without spoiling global mass fits~\cite{Bulgac:2018}.
For this reason, we start with a parameterization of pure neutron matter, then use the properties of symmetric nuclear matter to constrain the extrapolation in the proton fraction $x_p$.

To describe pure neutron matter we use a double polytrope for the energy per particle:
\begin{gather}
  E_n(n_n) = \frac{\E_n(n_n)}{n_n} =
  m_Nc^2
  + a\left(\frac{n_n}{\bar{n}_0}\right)^{\alpha}
  + b\left(\frac{n_n}{\bar{n}_0}\right)^{\beta},\nonumber\\
  P_n = n_n\left[
    a\alpha \left(\frac{n_n}{\bar{n}_0}\right)^{\alpha}
    + b\beta \left(\frac{n_n}{\bar{n}_0}\right)^{\beta}
    \right],\label{eq:Polytrope}
\end{gather}
where $m_N$ is the nucleon mass, $\bar{n}_0 = \SI{0.16}{fm^{-3}}$ is a constant (approximately the nuclear saturation density), and $a$, $b$, $\alpha$, and $\beta$ are four \gls{EoS} parameters.
This form was found to accurately fit \gls{QMC} calculations of the \gls{EoS} using nuclear Hamiltonians with realistic two-\ and three-body forces~\cite{Gandolfi:2009, *Gandolfi:2010b, *Gandolfi:2012, *Gandolfi:2014a}, and is consistent with recent \gls{QMC} results based on chiral \gls{EFT} interactions~\cite{Wlazlowski:2014a, Gandolfi:2014, Gandolfi:2015, Lynn:2015}.
For small proton fractions $\xp = n_p/(n_n + n_p)$, we perform an expansion:
\begin{multline}
  E_{np}(n_n, n_p) = 
  (1-\xp)E_n(n_n) + \xp\Bigl(m_pc^2 + \Sigma^p(n_B)\Bigr) +\\
  + \frac{\hbar^2(3\pi^2)^{2/3}}{2m^*}\xp^{5/3}n_B^{2/3} + \xp^2f_2(n_B) + \xp^3f_3(n_B) + \cdots
\end{multline}
where $m^*$ is the proton effective mass, and $\Sigma^p(n_B)$ describes the self-energy of the proton polaron.
This function is presently poorly constrained by \gls{QMC} and experimental data and all known results are consistent with a simple two-parameter quadratic expansion:
\begin{gather}
  \Sigma^p(n_B) = \mu_p\frac{n_B}{\bar{n}_0}\frac{u_p -
    \frac{n_B}{\bar{n}_0}}{u_p-1}
\end{gather}
where $\mu_p = \Sigma^p(\bar{n}_0)$ and $u_p = n_B/\bar{n}_0$ where $\Sigma^p(n_B)=0$ returns to zero.
(We expect $\Sigma^p(n_B)$ to curve up for higher densities due to the repulsive nature of nuclear three-body interactions).

The additional powers $f_n(n_B)$ are chosen to match the properties of nuclear matter to quadratic order in the isospin asymmetry $\beta_p$ and expansion away from saturation $\delta_n$:
\begin{gather}
  E^\text{sym}_{np}(n_n, n_p) = \varepsilon_0 + \frac{K_0}{2}\delta_n^2 +
  \Bigl(S_2 + L_2 \delta_n + \frac{K_2}{2}\delta_n^2\Bigr)\beta_p^2,\nonumber\\
  \beta_p = \frac{n_n-n_p}{n_B}, \qquad
   \delta_n = \frac{n_B - n_0}{3n_0}.
   \label{eq:E_np}
\end{gather}
Fitting two even powers, $\beta_p^0$ and $\beta_p^2$, and the lack of
odd powers uniquely defines the functions $f_2(n_B)$ through
$f_5(n_B)$, completing our characterization of the nuclear equation of
state in terms of the nuclear saturation density $n_0$, energy
$\varepsilon_0$, and incompressibility $K_0$; the symmetry energy
$S_2$, slope $L_2$ and incompressibility $K_2$. Note that a term proportional to $\beta_p^4$ is allowed in \cref{eq:E_np}, but our \gls{EoS} is unconstrained by this term, i.e., does not rely on the parabolic approximation \cref{eq:E_np}.

\subsection{Speed of Sound Parameterization of the Inner Core}\label{sec:cs}
Above densities $n_c \sim 2n_0$ the \gls{EoS} is virtually unconstrained.
The typical approximation at high density is in terms of a polytrope, but we choose a more physically motivated high-density \gls{EoS} parameterized in terms of the square of the speed of sound: $C(\E) = c_s^2(\E)/c^2 = P'(\E) \leq 1$ which approaches the \gls{PQCD} result $C(\E) \rightarrow 1/3$ at asymptotic densities.
Although the form of the function $C(\E)$ is unknown at finite density, its qualitative form at finite temperature suggests that it may first peak before returning to the asymptotic value~\cite{Alford:2016pc, Tews:2018kmu}.
We thus include a simple parameterization $C(\E)$ as a quadratic polynomial smoothly connecting to the homogeneous equation of state at a fixed transition energy density $\E_c$ reaching a maximum $C_{\text{max}} \leq 1$ at an energy density $\E_{\text{max}}$, then returning to $C = 1/3$ at which it remains for higher densities.
This core \gls{EoS} thus introduces three parameters $\E_c$, $C_{\text{max}}$, and $\E_{\text{max}}$.
To better understand the sensitivity of our results to the properties of the core, we include one slightly different form \gls{Central2c_trans} which has a first-order phase transition with discontinuity $\mathcal{E}_{\mathrm{trans}}$ at $\mathcal{E}_{c}$.

\subsection{Parameters}
Our equation of state is thus characterized by 18 parameters: $\sigma_\delta$ and $\mathcal{C}$, (\gls{CLDM}), $n_0$, $\varepsilon_0$, $K_0$, (symmetric nuclear matter), $S_2$, $L_2$, $K_2$, (symmetry energy), $a$, $\alpha$, $b$, $\beta$, (neutron matter) $\mu_p$, $u_p$, $m^*$, (proton polaron), and $\E_c$, $\E_{\text{max}}$, $C_{\text{max}}$ (core).
We explore various ranges of these parameters centered about the values listed in \cref{tab:Central2c}, which defines our base \gls{Central2c} \gls{EoS} model.
In addition to these central values, we repeat our analysis at a handful of different parameter values, defining the models listed in \cref{tab:eos}.
Some of these are referred to in the text, but a complete comparison is present in the supplement~\cite{EPAPS}.
We now discuss how these constraints are derived from gravitational wave observations.

\section{Gravitational Waveform}\label{sec:GW}

Gravitational waves from merging binary neutron star systems carry information about the nuclear equations of state. During late stages of inspiral tidal interactions between neutron stars can leave imprints on the \gls{GW} signal that is otherwise dominated by point-mass contributions.
As mentioned earlier, tidal responses of neutron stars can be quantified by the dimensionless tidal deformability parameter $\Lambda = \tfrac{2}{3} k_2 c^{10}R^5/(G M)^5$, where the second Love number $k_2$ is weakly sensitive to the matter distribution inside the star~\cite{Flanagan:2007ix}. The strong dependence of $\Lambda$ on the radius $R$ of neutron star allows us to extract information regarding nuclear \gls{EoS}.
Indeed, \gls{pN} theory is able to quantitatively describe the effect of the \gls{NS} \gls{EoS} on the signal by parameterizing the waveform in terms of $M$ and $\Lambda$ of component stars~\cite{Flanagan:2007ix, Vines:2011ud}.

Gravitational wave observations of inspiraling compact binaries involving neutron stars can therefore constrain $\Lambda$~\cite{TheLIGOScientific:2017qsa,Abbott:2018exr}. 
However, since the constraint on $\Lambda$ from a single \gls{BNS}  is weak for small to medium \gls{SNR} events, multiple observations of such systems will be required for remote sources to reduce the statistical error in $M$s and $\Lambda$s in order to discern the effects of similar \gls{EoS}~\cite{DelPozzo:2013PRL, Agathos:2015uaa,Bose:2017jvk}. 
Fortunately, tens-to-hundreds of binaries of this type~\cite{LIGOScientific:2018mvr} are expected to be observed over the next several years by the advanced (or ``second generation'') \gls{LIGO}.

We consider only non-spinning neutron stars here because astrophysically their spins $J$ are expected to be small when in a \gls{BNS} system; in particular it is believed that the dimensionless spin parameter $Jc/(GM^2) \leq 0.04$~\cite{Stovall:2018ouw,TheLIGOScientific:2017qsa}
We plan to study the effect of spin in a future follow up study.

The \gls{GW} signal from a \gls{BNS} system in a detector can be expressed as the strain
\begin{equation}
  h(t)= A(t)e^{i \Psi(t)}\,,
\end{equation}
where $A(t)$ and $\Psi(t)$ denote its amplitude and phase in the time domain.
For \gls{FIM}-based parameter estimation, we work with the Fourier transform $\tilde{h}$ of the strain above.
This is constructed by adding to the point-particle part of the \texttt{TaylorF2} model at 3.5\gls{pN}~\cite{Buonanno:2009zt}, a phase correction that is taken here to be the Fourier domain tidal waveform, with Pad\'{e} fits, as prescribed in Dietrich et al.~\cite{Dietrich:2017aum}.

\section{Population Models}
\label{sec:population-models}
We employ different sets of stellar evolution model parameters of \gls{ZAMS} binary stars each of which would lead to a binary neutron star system that merges within Hubble time.
The differences among stellar evolution models can be large, resulting in appreciable variation in the component mass distribution.
Since the tidal deformability parameter is sensitive to the masses, we explore four cases of mass distributions produced by population synthesis studies~\cite{Dominik:2012kk}.
These are more realistic than the uniform or Gaussian distributions owing to the application of stellar evolution mechanism of binary stars including two important factors, namely, metallicity and the nature of the common envelop interaction in the binary.

Metallicity plays the most dominant role in determining the strength of stellar winds in main sequence stars.
The larger the metallicity the larger the stellar winds, due to increased scattering cross-section of the electrons.
This results in increased mass loss; therefore, the remnant mass left behind at the end of main sequence phase is reduced.
This decreases the total baryonic mass content of the supernova engine at the onset of the explosion.
In our study, we consider two different variants of metallicities produced by~\cite{Dominik:2012kk}.
In the first case, the stellar evolution model was used with metallicity abundances being the same as solar metallicity, while in the second case 1/10th of solar metallicity was used.
The latter is termed to be of sub-solar metallicity. Component masses are narrowly peaked for solar metallicity systems while subsolar metalicity system produce a wider mass distribution.

\begin{figure}[tbp]
  \includegraphics[width=\columnwidth]{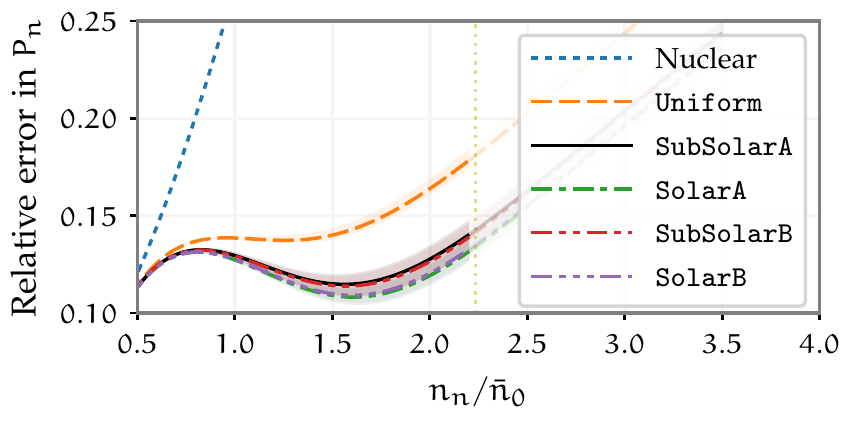}
  \caption{\label{fig:pc_pop}\glsreset{EoS}%
    (color online) Population model sensitivity of the constraint on the pressure of neutron matter $P_n(n_n)$ from $N_{\mathrm{obs}} = 15$ simulated merger events drawn from various different population models.
    The weaker constraints from the \gls{Uniform} model result from distributing the events over larger mass objects.
    As shown in \cref{fig:PCA_uniform}, this provides more information about the properties of the core at the expense of information about the lower-density regions that constrain neutron matter.
  }
\end{figure}

The second most important effect that can change the component masses of \gls{BNS} systems is the way mass transfer takes place during the common envelop phase of stellar evolution of the binary stars.
The mass transfer in the common envelop stage depends on the evolutionary phase of the two stars.
In one extreme case, for example, if the common envelop phase is initiated by the star in the Hertzsprung gap stage, it is likely to transfer a significant amount of orbital angular momentum to the entire binary system.
This case is denoted by ``submodel A'' in~\cite{Dominik:2012kk}.
On the other hand, depending on the nature of interaction between the core and the envelop, one possible outcome is that during each common envelop stage for Hertzsprung gap donor stars the outer envelope acquires the significant part of the orbital angular momentum and gets ejected from the system, leaving behind the cores of the two stars to inspiral.
This case is denoted by ``submodel B'' in~\cite{Dominik:2012kk}.
Furthermore, a higher metallicity in the parent star can result in greater mass loss and consequently a less massive remnant.
Therefore, we employ \gls{NS} populations resulting from solar metallicity stars as well as those with 10\% of solar metallicity.
These different characteristics lead to the following four categories of population models studied here:
\glsreset{StdASolar}
\glsreset{CompleteStdAsubsolarNSNS}
\glsreset{CompleteStdBsolarNSNS}
\glsreset{CompleteStdBsubsolarNSNS}
\glsreset{Uniform}
\begin{description}
\item[\Gls{StdASolar}]
  These are binary \gls{NS} populations produced by solar metallicity stars of the submodel A type.
\item[\Gls{CompleteStdAsubsolarNSNS}]
  These are binary \gls{NS} populations produced by sub-solar metallicity stars of the submodel A type.
\item[\Gls{CompleteStdBsolarNSNS}] 
  These are binary \gls{NS} populations produced by solar metallicity stars of the submodel B type.
\item[\Gls{CompleteStdBsubsolarNSNS}]
  These are binary \gls{NS} populations produced by sub-solar metallicity stars of the submodel B type.
\item[\Gls{Uniform}]
  Uniform sampling of neutron stars with masses between $1.2M_\odot$ and $1.8M_\odot$.
\end{description}

\begin{figure}[tb]
  \includegraphics[width=\columnwidth]{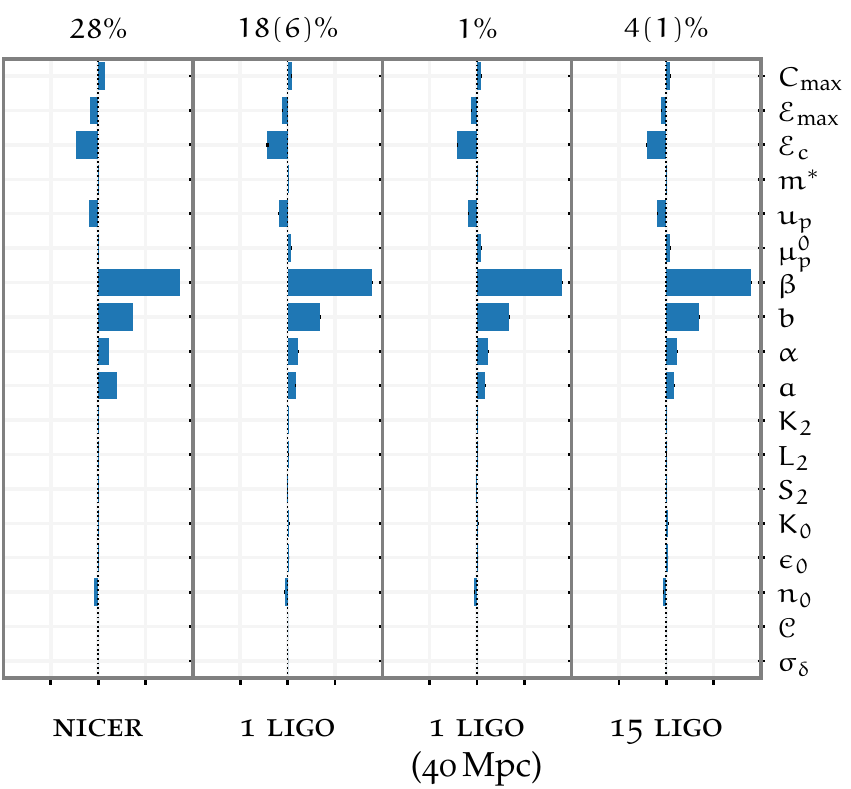}
  \caption{\label{fig:PCA}%
    (color online) Principal component analysis of the simulated observational data in terms of the \gls{EoS} parameters.
    Each column is a plot of the components of most significantly constrained eigenvector for the particular combination of observations listed at the bottom.
    These should be interpreted as follows: A linear combination of the log of the corresponding parameters is constrained to the tolerance shown at the top.
    The rightmost column shows the principal component analysis for $N_{\text{obs}}=15$ simulated merger events drawn from the \Gls{CompleteStdAsubsolarNSNS} population model, and is the same as the leftmost column of \cref{fig:PCA_all}.
    The $1\sigma$ errors in the tolerances, shown as small black strips in middle of the component bars, are obtained by performing 200 independent samples and demonstrate variation within the population model.
    \emph{(These errors are small here, but quite visible in the second principal components of \cref{fig:PCA_uniform}.)} 
  }
\end{figure}

\begin{figure}[tb]
  \includegraphics[width=\columnwidth]{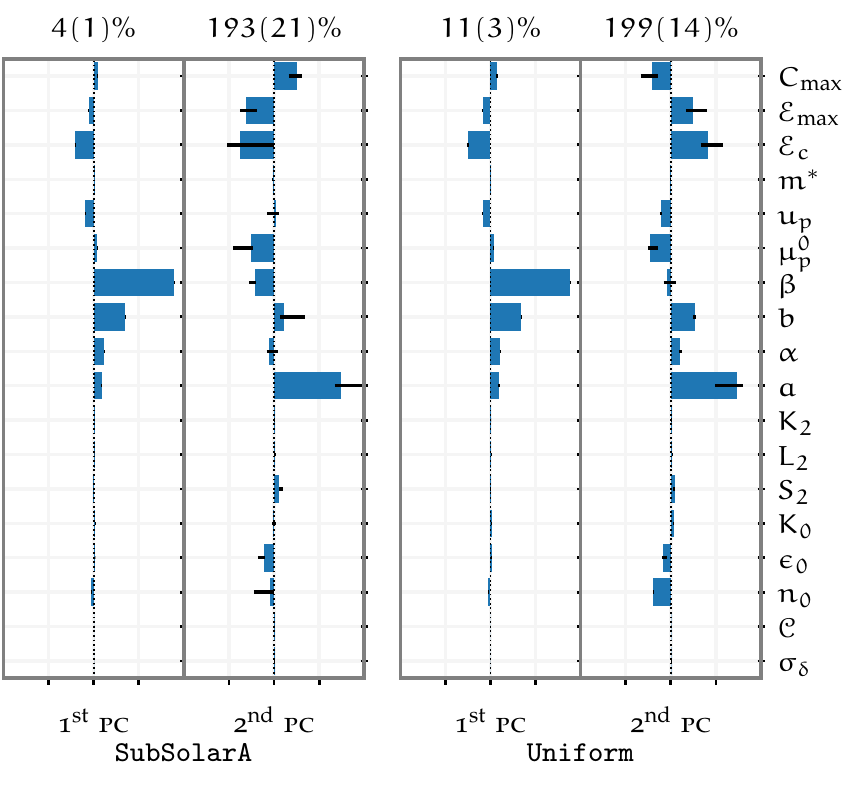}
  \caption{\label{fig:PCA_uniform}%
    (color online) First two principal components for $N_{\text{obs}}=15$ \gls{GW} observations drawn from the \Gls{CompleteStdAsubsolarNSNS} (left) and \Gls{Uniform} (right) population models.
    This demonstrates the wider distribution of masses in the \gls{Uniform} model as compared to \Gls{CompleteStdAsubsolarNSNS}.
    The narrow distribution in \Gls{CompleteStdAsubsolarNSNS} leads to tighter statistical constraints on the 1st principal component, but leaves other directions in parameter space poorly explored.
    In contrast, the \gls{Uniform} model distributes the 15 events over a larger range of masses, reducing the constraints on the 1st principle component, but providing more information about a second direction.
    Even for $N_{\text{obs}}=15$ observations, the next principal is poorly constrained at a level worse than 200\%: more observations would be required to constrain this component at a useful level.
    Thus, neutron-star observables seem to provide tight constraints in a single direction of parameter space.
  }
\end{figure}

\begin{figure*}[t]
  \includegraphics[width=\textwidth]{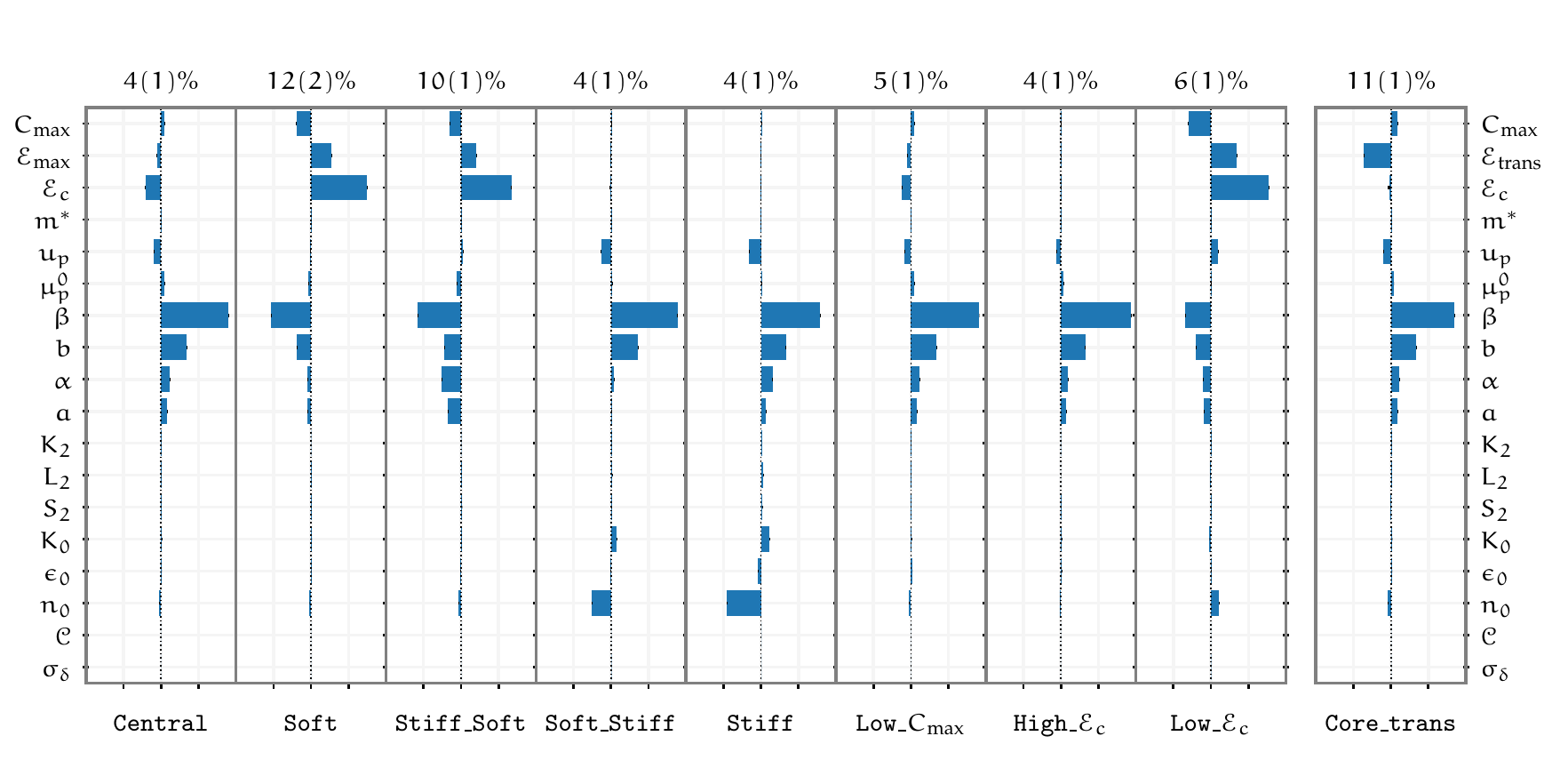}
  \caption{\label{fig:PCA_all}%
    (color online) Principal component analysis of $N_{\text{obs}}=15$ simulated merger events drawn from the \gls{CompleteStdAsubsolarNSNS} population model for each of the \gls{EoS} parameters listed in \cref{tab:eos}.
    The leftmost column thus corresponds to the rightmost column of \cref{fig:PCA}.
    This analysis makes clear the non-linear dependence of the problem on the \gls{EoS} parameters: neutron star observations constrain either properties of the core for parameter values such as \gls{Soft2c}, \gls{Stiff2c_Soft}, \gls{Central2c_low_Ec}, or \gls{Central2c_trans}, or the neutron matter \gls{EoS} for more central values.
    \exclude{(Note: The enhanced sensitivity of the several of the models to the saturation density $n_0$ and $K_0$ corresponds to stiffer equations of state where the tidal deformability is more influenced by the crust.)}
  }
\end{figure*}

\section{Statistics and Methods}
\label{sec:stat}

\newcommand{\vth}{\boldsymbol{\vartheta}} Given a particular parameterization of the \gls{EoS}, we compute the mass $M$, radius $R$, and tidal deformability parameter $\Lambda$ of a \gls{NS} with a given central density by solving the \gls{TOV} equations (see e.g.~\cite{Hinderer:2007mb, Postnikov:2010}).
The signals (gravitational waveforms) from merging neutron stars is computed with the numerical relativity based frequency-domain model~\cite{Dietrich:2017aum} mentioned above.
From those waveforms, we compute the corresponding \gls{FIM} characterizing the correlated uncertainties of the masses, $M_1$ and $M_2$, and the tidal deformabilities, $\Lambda_1$ and $\Lambda_2$ (maximizing the matched-filter over the source distance, signal time, and phase at colaescence~\cite{Ajith:2009fz}), to estimate the information obtainable in a merger event at \gls{aLIGO} design sensitivity, as described below.

\paragraph{Statistical Analysis}
To estimate how large the noise-limited errors are of the \gls{BNS} parameters $\vth$, we begin by modeling the measured values after the \glspl{MLE}~\cite{Helstrom}.
Owing to noise, the \gls{MLE} will fluctuate about the respective true values, i.e., $\hat{\vth} = \vth + \delta\vth$, where $\delta\vth$ is the random error.
The extent of these fluctuations is estimated by the elements of the variance-covariance matrix, $\gamma^{ab}=\overline{\delta\vartheta^a \delta\vartheta^b}$~\cite{Helstrom}.

The matrix $\gamma^{ab}$ is bounded by the signal via the Cramer-Rao inequality, which states that
\begin{gather}
  \norm{\mat{\gamma}} \geq \norm{\mat{\Gamma}}^{-1}\;,
\end{gather}
where $\mat{\Gamma}$ is the \gls{FIM}:
\begin{align}
  \label{eq:Fisher}
  \Gamma_{ab}
   & = \braket{\partial_a \tilde{h}(\vth), \partial_b \tilde{h}(\vth)} \nonumber                                 \\
   & \equiv  4 \Re \int \d{f}\;\frac{\partial_a \tilde{h}^{*}(f; \vth)\;\partial_b \tilde{h}(f; \vth)}{S_{h}(f)}\,.
\end{align}
Above, $\partial_a$ is the partial derivative with respect to the parameter $\vartheta^a$ and $S_{h}(f)$ is the one-sided noise \gls{PSD}~\cite{Helstrom}.
We take the latter to be the zero-detuned high-power \gls{PSD} for \gls{aLIGO}~\cite{aLIGO_ZDHP}.
Therefore, $\Delta \vartheta^a \equiv \left(~\overline{\delta \vartheta^a\,\delta \vartheta^a}~\right)^{1/2} = \Gamma_{aa}^{-1/2}$ gives the lower bound on the \gls{rms} error in estimating $\vartheta^a$.
The two are equal in the limit of large \gls{SNR} (see, e.g.,~\cite{Vallisneri:2007ev}).
The error estimates listed here are the $\Delta\vartheta^a$ obtained from the \gls{FIM}.

The \gls{FIM} method is known to underestimate the error in the estimation of the masses~\cite{Rodriguez:2013mla}.
We therefore used error-estimates for total-mass $M_{\mathrm{tot}}$ and mass-ratio $q$ (i.e., the ratio of the lighter mass to heavier mass) that were obtained with Bayesian methods in Ref.~\cite{Rodriguez:2013oaa}, and set them such that the $1\sigma$ error is $\Delta M_{\mathrm{tot}} / M_{\mathrm{tot}} = 2\%$ and $\Delta q = 0.28$, respectively, at a single-detector \gls{SNR} of 10.

The corresponding error in $\Lambda$ for individual systems is consistent with that found in the available literature~\cite{Damour:2012yf, Lackey:2011vz, Agathos:2015uaa}.
While these studies probe how accurately $\Lambda$ can be measured from \gls{GW} observations, they do not explore the effect of directly including inputs from nuclear theory, which is the point of this work.

To translate these correlated uncertainties in observables $M$s and $\Lambda$s (assuming effects of component spins to be small for $Jc/(GM^2) \leq 0.04$) to nuclear physics parameters, the \gls{FIM} generated from the waveforms described above is transformed to the space of nuclear parameters $\Theta$ via the Jacobian $\partial \theta/\partial \Theta$ such as the partial derivative $\partial M/\partial \alpha$.
These are then combined with a \gls{FIM} from the base nuclear uncertainties, and information about neutron star masses and radii at levels expected of \gls{NASA}'s \gls{NICER} mission to obtain a final covariance matrix for the 18 parameters.

The Fisher method for estimating errors has limitations, one of the main being the need for a high \gls{SNR}.
Bayesian methods are more reliable, but computationally much more expensive.
For this latter reason use Fisher methods, whose computationally efficiency allows us to reduce source selection effects on the error estimates.
We are able to quickly compute the \gls{FIM} for hundreds of binaries, characterizing the variance within the population models.
In spite of the drawbacks, the Fisher errors quoted here make the case to invest in Bayesian methods.

\paragraph{Methodology}
For a given population synthesis model, we simulate ten thousand \gls{BNS} systems and distribute them uniformly in comoving volume between a luminosity distance of \SI{100}{Mpc} and \SI{400}{Mpc}.
The latter limit is not too far from the horizon distance ($\sim\SI{450}{Mpc}$) of the network of \gls{aLIGO} and Advanced Virgo detectors beyond which \gls{BNS} sources will produce signals with network \gls{SNR} of less than 8.
Also, below \SI{100}{Mpc} we expect almost an order of magnitude fewer sources than those up to a distance of \SI{200}{Mpc}. This fact notwithstanding the measurement precision for a nearby source (\gls{GW170817} was at a distance of $\sim$\SI{40}{Mpc}) can rival that of a population of more distant sources. This is why we also present results for a \gls{GW170817}-like source at aLIGO design sensitivity.

Our main results are summarized in \cref{fig:progressive_constraints}, which shows how the constraints on the pressure of pure neutron matter $P_n(n_n)$ improve as a function of additional \gls{NICER} or \gls{LIGO} observations.
We start from the errors listed in \cref{tab:Central2c} which, for the purposes of this analysis, we interpret as uncorrelated $1\sigma$ normal errors for the parameters.
In general, errors have been over-estimated to ensure that our results are conservative.
The resulting \gls{FIM} -- a diagonal matrix of the inverse variances $\sigma_p^{-2}$ -- provides our starting point.
From this \gls{FIM}, we use forward error propagation to determine the error in pressure which we label ``Nuclear''.

The largest uncertainty comes from the form of the \gls{EoS} in the core of the neutron star.
Although a description in terms of homogeneous nuclear matter may persist to some depth, it is likely that there is some sort of phase transition to hyperonic or strange quark matter.
The core \gls{EoS} is thus largely unknown.
To assess the impact of large variations in the core \gls{EoS}, we compare the constraints obtained under a rather large variation of the core parameters, as well as in the presence of a strong first-order phase transition (\gls{Central2c_trans}).
This comparison was summarized in \cref{fig:core_constraints}.
Here we see rather large sensitivity to a smaller $\mathcal{E}_c$ as expected:
if this is small, the core transition occurs at low density, and not enough conventional nuclear matter exists to be sensitive to \gls{GW} observations.
As long as the core transition is above $2n_0$ or so, the constraints on $P_n$ are relatively insensitive to the form of the core \gls{EoS} unless there is a strong first-order phase transition.

\glsreset{EoS}
\section{Conclusion}
The \gls{GW170817} event demonstrated that useful constraints on the neutron star structure can be obtained from \glspl{GW}.
In this article we have addressed how future observations can provide more detailed constraints on the properties of dense matter.
By separating the neutron star into four distinct regions, and providing a unique nuclear physics based parameterization of the \gls{EoS} of the crust and outer-core, we have analyzed how measurements of the tidal deformability can constrain nuclear properties of dense matter.
Our parameterization, which uses the same underlying \gls{EoS} of neutron matter in both inner crust and outer core, allowed us to estimate for the first time constraints on the \gls{EoS} of pure neutron matter in the density interval where controlled calculations are becoming feasible.
These constraints, as they become available, will provide valuable guidance for nuclear physics.
In the inner core, where the \gls{EoS} is poorly constrained, the speed of sound is allowed to vary over a large range constrained only by causality and the requirement that the \gls{EoS} produce a 2 solar mass neutron star.
We have taken first steps to study how the large uncertainties associated with the \gls{EoS} of the inner core limits our ability to constrain the \gls{EoS} of neutron matter in the outer core.
The results we obtain suggest that, in the absence of strong first-order transitions in the core, even a handful of detections can constrain the pressure of neutron matter in the density interval between $n_0$ and $2n_0$ to better than 20\%.

The principal component analysis presented in \cref{fig:PCA} suggests that future \gls{LIGO} observations will provide strong constraints on the density dependence of the pure neutron matter \gls{EoS} in the outer core.
In particular, we find that the exponent $\beta$ in the neutron matter \gls{EoS} defined in \cref{eq:Polytrope} will be well constrained.
As expected, the nuclear physics parameters are better constrained when the outer core makes the dominant contribution to the tidal deformability.
This is the case when the neutron matter \gls{EoS} is stiff in the dense regions of the outer core and for low-mass neutron stars.
If instead, the \gls{EoS} in the outer core is soft or if a strong first-order phase transition were to occur at relatively low-density, constraints on the neutron matter \gls{EoS} are weaker.
In these cases, the inner core has a larger impact on the tidal deformability and \gls{GW} detections will provide constraints for matter encountered in the inner core.

Although our focus here was to study the impact of the most common events that occur at large distances, we find that a single close by event similar to \gls{GW170817} at \SI{40}{Mpc} at design sensitivity will provide valuable constraints.
However, in the absence of such a nearby event, similar constraints may be realized by a dozen or so more distant events.

One limitation of our study is the simple parameterization of the \gls{EoS} of the inner core.
While this is adequate as a first step, to constrain the \gls{EoS} of the inner core, a parameterization that allows for larger variability at high density will be needed.
In addition, to gain more confidence in the constraints we have presented for neutron matter, it will be necessary to systematically marginalize over population models for neutron star masses and spins, and the uncertainty in the \gls{EoS} of the inner core.
A Bayesian approach would be better suited for this purpose, and we are in the processes of developing computer programs needed for such a study.

\begin{acknowledgments}
  We thank K.~G.~Arun for helpful discussions and early collaboration on waveform models with tidal corrections.
  We also thank Philippe Landry for carefully reading the manuscript and making useful comments.
  \mysc{Sb} acknowledges partial support from the Navajbai Ratan Tata Trust.
  \mysc{Sr} acknowledges support from the \mysc{us} Department of Energy Grant No.~\mysc{de-fg02-00er41132} and from the National Science Foundation Grant No.~\mysc{phy-1430152} (\mysc{jina} Center for the Evolution of the Elements).
  \mysc{Am} acknowledges partial support from the \mysc{serb} Start-Up Research for Young Scientists Scheme project Grant No.~\mysc{sb/ftp/ps-067/2014}, \mysc{dst}, India.
\end{acknowledgments}

\providecommand{\selectlanguage}[1]{}
\renewcommand{\selectlanguage}[1]{}
\bibliography{master,local}

\begin{thebibliography}{74}%
\makeatletter
\providecommand \@ifxundefined [1]{%
 \@ifx{#1\undefined}
}%
\providecommand \@ifnum [1]{%
 \ifnum #1\expandafter \@firstoftwo
 \else \expandafter \@secondoftwo
 \fi
}%
\providecommand \@ifx [1]{%
 \ifx #1\expandafter \@firstoftwo
 \else \expandafter \@secondoftwo
 \fi
}%
\providecommand \natexlab [1]{#1}%
\providecommand \enquote  [1]{``#1''}%
\providecommand \bibnamefont  [1]{#1}%
\providecommand \bibfnamefont [1]{#1}%
\providecommand \citenamefont [1]{#1}%
\providecommand \href@noop [0]{\@secondoftwo}%
\providecommand \href [0]{\begingroup \@sanitize@url \@href}%
\providecommand \@href[1]{\@@startlink{#1}\@@href}%
\providecommand \@@href[1]{\endgroup#1\@@endlink}%
\providecommand \@sanitize@url [0]{\catcode `\\12\catcode `\$12\catcode
  `\&12\catcode `\#12\catcode `\^12\catcode `\_12\catcode `\%12\relax}%
\providecommand \@@startlink[1]{}%
\providecommand \@@endlink[0]{}%
\providecommand \url  [0]{\begingroup\@sanitize@url \@url }%
\providecommand \@url [1]{\endgroup\@href {#1}{\urlprefix }}%
\providecommand \urlprefix  [0]{URL }%
\providecommand \Eprint [0]{\href }%
\providecommand \doibase [0]{http://dx.doi.org/}%
\providecommand \selectlanguage [0]{\@gobble}%
\providecommand \bibinfo  [0]{\@secondoftwo}%
\providecommand \bibfield  [0]{\@secondoftwo}%
\providecommand \translation [1]{[#1]}%
\providecommand \BibitemOpen [0]{}%
\providecommand \bibitemStop [0]{}%
\providecommand \bibitemNoStop [0]{.\EOS\space}%
\providecommand \EOS [0]{\spacefactor3000\relax}%
\providecommand \BibitemShut  [1]{\csname bibitem#1\endcsname}%
\let\auto@bib@innerbib\@empty
\bibitem [{\citenamefont {Aasi}\ \emph {et~al.}(2015)\citenamefont {Aasi} \emph
  {et~al.}}]{TheLIGOScientific:2014jea}%
  \BibitemOpen
  \bibfield  {author} {\bibinfo {author} {\bibfnamefont {J.}~\bibnamefont
  {Aasi}} \emph {et~al.} (\bibinfo {collaboration} {LIGO Scientific}),\ }\href
  {\doibase 10.1088/0264-9381/32/7/074001} {\bibfield  {journal} {\bibinfo
  {journal} {Class. Quant. Grav.}\ }\textbf {\bibinfo {volume} {32}},\ \bibinfo
  {pages} {074001} (\bibinfo {year} {2015})},\ \Eprint
  {http://arxiv.org/abs/1411.4547} {arXiv:1411.4547 [gr-qc]} \BibitemShut
  {NoStop}%
\bibitem [{\citenamefont {Acernese}\ \emph {et~al.}(2015)\citenamefont
  {Acernese} \emph {et~al.}}]{TheVirgo:2014hva}%
  \BibitemOpen
  \bibfield  {author} {\bibinfo {author} {\bibfnamefont {F.}~\bibnamefont
  {Acernese}} \emph {et~al.} (\bibinfo {collaboration} {VIRGO}),\ }\href
  {\doibase 10.1088/0264-9381/32/2/024001} {\bibfield  {journal} {\bibinfo
  {journal} {Class. Quant. Grav.}\ }\textbf {\bibinfo {volume} {32}},\ \bibinfo
  {pages} {024001} (\bibinfo {year} {2015})},\ \Eprint
  {http://arxiv.org/abs/1408.3978} {arXiv:1408.3978 [gr-qc]} \BibitemShut
  {NoStop}%
\bibitem [{\citenamefont {Abbott}\ \emph
  {et~al.}(2017{\natexlab{a}})\citenamefont {Abbott} \emph
  {et~al.}}]{TheLIGOScientific:2017qsa}%
  \BibitemOpen
  \bibfield  {author} {\bibinfo {author} {\bibfnamefont {B.~P.}\ \bibnamefont
  {Abbott}} \emph {et~al.} (\bibinfo {collaboration} {LIGO Scientific,
  Virgo}),\ }\href {\doibase 10.1103/PhysRevLett.119.161101} {\bibfield
  {journal} {\bibinfo  {journal} {Phys. Rev. Lett.}\ }\textbf {\bibinfo
  {volume} {119}},\ \bibinfo {pages} {161101} (\bibinfo {year}
  {2017}{\natexlab{a}})},\ \Eprint {http://arxiv.org/abs/1710.05832}
  {arXiv:1710.05832 [gr-qc]} \BibitemShut {NoStop}%
\bibitem [{\citenamefont {Abbott}\ \emph
  {et~al.}(2017{\natexlab{b}})\citenamefont {Abbott} \emph
  {et~al.}}]{GBM:2017lvd}%
  \BibitemOpen
  \bibfield  {author} {\bibinfo {author} {\bibfnamefont {B.~P.}\ \bibnamefont
  {Abbott}} \emph {et~al.} (\bibinfo {collaboration} {LIGO Scientific, Virgo,
  Fermi GBM, INTEGRAL, IceCube, AstroSat Cadmium Zinc Telluride Imager Team,
  IPN, Insight-Hxmt, ANTARES, Swift, AGILE Team, 1M2H Team, Dark Energy Camera
  GW-EM, DES, DLT40, GRAWITA, Fermi-LAT, ATCA, ASKAP, Las Cumbres Observatory
  Group, OzGrav, DWF (Deeper Wider Faster Program), AST3, CAASTRO, VINROUGE,
  MASTER, J-GEM, GROWTH, JAGWAR, CaltechNRAO, TTU-NRAO, NuSTAR, Pan-STARRS,
  MAXI Team, TZAC Consortium, KU, Nordic Optical Telescope, ePESSTO, GROND,
  Texas Tech University, SALT Group, TOROS, BOOTES, MWA, CALET, IKI-GW
  Follow-up, H.E.S.S., LOFAR, LWA, HAWC, Pierre Auger, ALMA, Euro VLBI Team, Pi
  of Sky, Chandra Team at McGill University, DFN, ATLAS Telescopes, High Time
  Resolution Universe Survey, RIMAS, RATIR, SKA South Africa/MeerKAT}),\ }\href
  {\doibase 10.3847/2041-8213/aa91c9} {\bibfield  {journal} {\bibinfo
  {journal} {Astrophys. J.}\ }\textbf {\bibinfo {volume} {848}},\ \bibinfo
  {pages} {L12} (\bibinfo {year} {2017}{\natexlab{b}})},\ \Eprint
  {http://arxiv.org/abs/1710.05833} {arXiv:1710.05833 [astro-ph.HE]}
  \BibitemShut {NoStop}%
\bibitem [{\citenamefont {Abbott}\ \emph
  {et~al.}(2017{\natexlab{c}})\citenamefont {Abbott} \emph
  {et~al.}}]{Abbott:2017xzu}%
  \BibitemOpen
  \bibfield  {author} {\bibinfo {author} {\bibfnamefont {B.~P.}\ \bibnamefont
  {Abbott}} \emph {et~al.} (\bibinfo {collaboration} {LIGO Scientific, Virgo,
  1M2H, Dark Energy Camera GW-E, DES, DLT40, Las Cumbres Observatory, VINROUGE,
  MASTER}),\ }\href {\doibase 10.1038/nature24471} {\bibfield  {journal}
  {\bibinfo  {journal} {Nature}\ }\textbf {\bibinfo {volume} {551}},\ \bibinfo
  {pages} {85} (\bibinfo {year} {2017}{\natexlab{c}})},\ \Eprint
  {http://arxiv.org/abs/1710.05835} {arXiv:1710.05835 [astro-ph.CO]}
  \BibitemShut {NoStop}%
\bibitem [{\citenamefont {Chen}\ \emph {et~al.}(2018)\citenamefont {Chen},
  \citenamefont {Fishbach},\ and\ \citenamefont {Holz}}]{Chen:2017rfc}%
  \BibitemOpen
  \bibfield  {author} {\bibinfo {author} {\bibfnamefont {H.-Y.}\ \bibnamefont
  {Chen}}, \bibinfo {author} {\bibfnamefont {M.}~\bibnamefont {Fishbach}}, \
  and\ \bibinfo {author} {\bibfnamefont {D.~E.}\ \bibnamefont {Holz}},\ }\href
  {\doibase 10.1038/s41586-018-0606-0} {\bibfield  {journal} {\bibinfo
  {journal} {Nature}\ }\textbf {\bibinfo {volume} {562}},\ \bibinfo {pages}
  {545} (\bibinfo {year} {2018})},\ \Eprint {http://arxiv.org/abs/1712.06531}
  {arXiv:1712.06531 [astro-ph.CO]} \BibitemShut {NoStop}%
\bibitem [{\citenamefont {Nair}\ \emph {et~al.}(2018)\citenamefont {Nair},
  \citenamefont {Bose},\ and\ \citenamefont {Saini}}]{Nair:2018ign}%
  \BibitemOpen
  \bibfield  {author} {\bibinfo {author} {\bibfnamefont {R.}~\bibnamefont
  {Nair}}, \bibinfo {author} {\bibfnamefont {S.}~\bibnamefont {Bose}}, \ and\
  \bibinfo {author} {\bibfnamefont {T.~D.}\ \bibnamefont {Saini}},\ }\href
  {\doibase 10.1103/PhysRevD.98.023502} {\bibfield  {journal} {\bibinfo
  {journal} {Phys. Rev.}\ }\textbf {\bibinfo {volume} {D98}},\ \bibinfo {pages}
  {023502} (\bibinfo {year} {2018})},\ \Eprint
  {http://arxiv.org/abs/1804.06085} {arXiv:1804.06085 [astro-ph.CO]}
  \BibitemShut {NoStop}%
\bibitem [{\citenamefont {Soares-Santos}\ \emph {et~al.}(2019)\citenamefont
  {Soares-Santos} \emph {et~al.}}]{Soares-Santos:2019irc}%
  \BibitemOpen
  \bibfield  {author} {\bibinfo {author} {\bibfnamefont {M.}~\bibnamefont
  {Soares-Santos}} \emph {et~al.} (\bibinfo {collaboration} {DES, LIGO
  Scientific, Virgo}),\ }\href@noop {} {\bibfield  {journal} {\bibinfo
  {journal} {Submitted to: Astrophys. J.}\ } (\bibinfo {year} {2019})},\
  \Eprint {http://arxiv.org/abs/1901.01540} {arXiv:1901.01540 [astro-ph.CO]}
  \BibitemShut {NoStop}%
\bibitem [{\citenamefont {Abbott}\ \emph
  {et~al.}(2018{\natexlab{a}})\citenamefont {Abbott} \emph
  {et~al.}}]{Abbott:2018exr}%
  \BibitemOpen
  \bibfield  {author} {\bibinfo {author} {\bibfnamefont {B.~P.}\ \bibnamefont
  {Abbott}} \emph {et~al.} (\bibinfo {collaboration} {LIGO Scientific,
  Virgo}),\ }\href {\doibase 10.1103/PhysRevLett.121.161101} {\bibfield
  {journal} {\bibinfo  {journal} {Phys. Rev. Lett.}\ }\textbf {\bibinfo
  {volume} {121}},\ \bibinfo {pages} {161101} (\bibinfo {year}
  {2018}{\natexlab{a}})},\ \Eprint {http://arxiv.org/abs/1805.11581}
  {arXiv:1805.11581 [gr-qc]} \BibitemShut {NoStop}%
\bibitem [{\citenamefont {Flanagan}\ and\ \citenamefont
  {Hinderer}(2008)}]{Flanagan:2007ix}%
  \BibitemOpen
  \bibfield  {author} {\bibinfo {author} {\bibfnamefont {E.~E.}\ \bibnamefont
  {Flanagan}}\ and\ \bibinfo {author} {\bibfnamefont {T.}~\bibnamefont
  {Hinderer}},\ }\href {\doibase 10.1103/PhysRevD.77.021502} {\bibfield
  {journal} {\bibinfo  {journal} {Phys. Rev.}\ }\textbf {\bibinfo {volume}
  {D77}},\ \bibinfo {pages} {021502} (\bibinfo {year} {2008})},\ \Eprint
  {http://arxiv.org/abs/0709.1915} {arXiv:0709.1915 [astro-ph]} \BibitemShut
  {NoStop}%
\bibitem [{\citenamefont {De}\ \emph {et~al.}(2018)\citenamefont {De},
  \citenamefont {Finstad}, \citenamefont {Lattimer}, \citenamefont {Brown},
  \citenamefont {Berger},\ and\ \citenamefont {Biwer}}]{De:2018uhw}%
  \BibitemOpen
  \bibfield  {author} {\bibinfo {author} {\bibfnamefont {S.}~\bibnamefont
  {De}}, \bibinfo {author} {\bibfnamefont {D.}~\bibnamefont {Finstad}},
  \bibinfo {author} {\bibfnamefont {J.~M.}\ \bibnamefont {Lattimer}}, \bibinfo
  {author} {\bibfnamefont {D.~A.}\ \bibnamefont {Brown}}, \bibinfo {author}
  {\bibfnamefont {E.}~\bibnamefont {Berger}}, \ and\ \bibinfo {author}
  {\bibfnamefont {C.~M.}\ \bibnamefont {Biwer}},\ }\href@noop {} {\  (\bibinfo
  {year} {2018})},\ \Eprint {http://arxiv.org/abs/1804.08583} {arXiv:1804.08583
  [astro-ph.HE]} \BibitemShut {NoStop}%
\bibitem [{\citenamefont {Tews}\ \emph
  {et~al.}(2018{\natexlab{a}})\citenamefont {Tews}, \citenamefont {Margueron},\
  and\ \citenamefont {Reddy}}]{Tews:2018iwm}%
  \BibitemOpen
  \bibfield  {author} {\bibinfo {author} {\bibfnamefont {I.}~\bibnamefont
  {Tews}}, \bibinfo {author} {\bibfnamefont {J.}~\bibnamefont {Margueron}}, \
  and\ \bibinfo {author} {\bibfnamefont {S.}~\bibnamefont {Reddy}},\
  }\href@noop {} {\  (\bibinfo {year} {2018}{\natexlab{a}})},\ \Eprint
  {http://arxiv.org/abs/1804.02783} {arXiv:1804.02783 [nucl-th]} \BibitemShut
  {NoStop}%
\bibitem [{\citenamefont {Abbott}\ \emph {et~al.}(2019)\citenamefont {Abbott}
  \emph {et~al.}}]{Abbott:2018wiz}%
  \BibitemOpen
  \bibfield  {author} {\bibinfo {author} {\bibfnamefont {B.~P.}\ \bibnamefont
  {Abbott}} \emph {et~al.} (\bibinfo {collaboration} {LIGO Scientific,
  Virgo}),\ }\href {\doibase 10.1103/PhysRevX.9.011001} {\bibfield  {journal}
  {\bibinfo  {journal} {Phys. Rev.}\ }\textbf {\bibinfo {volume} {X9}},\
  \bibinfo {pages} {011001} (\bibinfo {year} {2019})},\ \Eprint
  {http://arxiv.org/abs/1805.11579} {arXiv:1805.11579 [gr-qc]} \BibitemShut
  {NoStop}%
\bibitem [{\citenamefont {Tews}\ \emph
  {et~al.}(2018{\natexlab{b}})\citenamefont {Tews}, \citenamefont {Carlson},
  \citenamefont {Gandolfi},\ and\ \citenamefont {Reddy}}]{Tews:2018kmu}%
  \BibitemOpen
  \bibfield  {author} {\bibinfo {author} {\bibfnamefont {I.}~\bibnamefont
  {Tews}}, \bibinfo {author} {\bibfnamefont {J.}~\bibnamefont {Carlson}},
  \bibinfo {author} {\bibfnamefont {S.}~\bibnamefont {Gandolfi}}, \ and\
  \bibinfo {author} {\bibfnamefont {S.}~\bibnamefont {Reddy}},\ }\href
  {\doibase 10.3847/1538-4357/aac267} {\bibfield  {journal} {\bibinfo
  {journal} {Astrophys. J.}\ }\textbf {\bibinfo {volume} {860}},\ \bibinfo
  {pages} {149} (\bibinfo {year} {2018}{\natexlab{b}})},\ \Eprint
  {http://arxiv.org/abs/1801.01923} {arXiv:1801.01923 [nucl-th]} \BibitemShut
  {NoStop}%
\bibitem [{\citenamefont {Gandolfi}\ \emph {et~al.}(2009)\citenamefont
  {Gandolfi}, \citenamefont {Illarionov}, \citenamefont {Schmidt},
  \citenamefont {Pederiva},\ and\ \citenamefont {Fantoni}}]{Gandolfi:2009}%
  \BibitemOpen
  \bibfield  {author} {\bibinfo {author} {\bibfnamefont {S.}~\bibnamefont
  {Gandolfi}}, \bibinfo {author} {\bibfnamefont {A.~Y.}\ \bibnamefont
  {Illarionov}}, \bibinfo {author} {\bibfnamefont {K.~E.}\ \bibnamefont
  {Schmidt}}, \bibinfo {author} {\bibfnamefont {F.}~\bibnamefont {Pederiva}}, \
  and\ \bibinfo {author} {\bibfnamefont {S.}~\bibnamefont {Fantoni}},\ }\href
  {\doibase 10.1103/PhysRevC.79.054005} {\bibfield  {journal} {\bibinfo
  {journal} {Phys. Rev. C}\ }\textbf {\bibinfo {volume} {79}},\ \bibinfo
  {pages} {054005} (\bibinfo {year} {2009})}\BibitemShut {NoStop}%
\bibitem [{\citenamefont {Gandolfi}\ \emph {et~al.}(2010)\citenamefont
  {Gandolfi}, \citenamefont {Illarionov}, \citenamefont {Fantoni},
  \citenamefont {Miller}, \citenamefont {Pederiva},\ and\ \citenamefont
  {Schmidt}}]{Gandolfi:2010b}%
  \BibitemOpen
  \bibfield  {author} {\bibinfo {author} {\bibfnamefont {S.}~\bibnamefont
  {Gandolfi}}, \bibinfo {author} {\bibfnamefont {A.~Y.}\ \bibnamefont
  {Illarionov}}, \bibinfo {author} {\bibfnamefont {S.}~\bibnamefont {Fantoni}},
  \bibinfo {author} {\bibfnamefont {J.~C.}\ \bibnamefont {Miller}}, \bibinfo
  {author} {\bibfnamefont {F.}~\bibnamefont {Pederiva}}, \ and\ \bibinfo
  {author} {\bibfnamefont {K.~E.}\ \bibnamefont {Schmidt}},\ }\href {\doibase
  10.1111/j.1745-3933.2010.00829.x} {\bibfield  {journal} {\bibinfo  {journal}
  {MNRAS}\ }\textbf {\bibinfo {volume} {404}},\ \bibinfo {pages} {L35}
  (\bibinfo {year} {2010})}\BibitemShut {NoStop}%
\bibitem [{\citenamefont {Gandolfi}\ \emph {et~al.}(2012)\citenamefont
  {Gandolfi}, \citenamefont {Carlson},\ and\ \citenamefont
  {Reddy}}]{Gandolfi:2012}%
  \BibitemOpen
  \bibfield  {author} {\bibinfo {author} {\bibfnamefont {S.}~\bibnamefont
  {Gandolfi}}, \bibinfo {author} {\bibfnamefont {J.}~\bibnamefont {Carlson}}, \
  and\ \bibinfo {author} {\bibfnamefont {S.}~\bibnamefont {Reddy}},\ }\href
  {\doibase 10.1103/PhysRevC.85.032801} {\bibfield  {journal} {\bibinfo
  {journal} {Phys. Rev. C}\ }\textbf {\bibinfo {volume} {85}},\ \bibinfo
  {pages} {032801} (\bibinfo {year} {2012})},\ \Eprint
  {http://arxiv.org/abs/1101.1921} {arXiv:1101.1921} \BibitemShut {NoStop}%
\bibitem [{\citenamefont {Gandolfi}\ \emph
  {et~al.}(2014{\natexlab{a}})\citenamefont {Gandolfi}, \citenamefont
  {Carlson}, \citenamefont {Reddy}, \citenamefont {Steiner},\ and\
  \citenamefont {Wiringa}}]{Gandolfi:2014a}%
  \BibitemOpen
  \bibfield  {author} {\bibinfo {author} {\bibfnamefont {S.}~\bibnamefont
  {Gandolfi}}, \bibinfo {author} {\bibfnamefont {J.}~\bibnamefont {Carlson}},
  \bibinfo {author} {\bibfnamefont {S.}~\bibnamefont {Reddy}}, \bibinfo
  {author} {\bibfnamefont {A.}~\bibnamefont {Steiner}}, \ and\ \bibinfo
  {author} {\bibfnamefont {R.}~\bibnamefont {Wiringa}},\ }\href {\doibase
  10.1140/epja/i2014-14010-5} {\bibfield  {journal} {\bibinfo  {journal} {Eur.
  Phys. J. A}\ }\textbf {\bibinfo {volume} {50}},\ \bibinfo {pages} {10}
  (\bibinfo {year} {2014}{\natexlab{a}})},\ \Eprint
  {http://arxiv.org/abs/1307.5815} {arXiv:1307.5815 [nucl-th]} \BibitemShut
  {NoStop}%
\bibitem [{\citenamefont {Miller}\ and\ \citenamefont
  {Lamb}(2016)}]{Miller:2016a}%
  \BibitemOpen
  \bibfield  {author} {\bibinfo {author} {\bibfnamefont {M.~C.}\ \bibnamefont
  {Miller}}\ and\ \bibinfo {author} {\bibfnamefont {F.~K.}\ \bibnamefont
  {Lamb}},\ }\href {\doibase 10.1140/epja/i2016-16063-8} {\bibfield  {journal}
  {\bibinfo  {journal} {Eur. J. Phys. A}\ }\textbf {\bibinfo {volume} {52}},\
  \bibinfo {pages} {63} (\bibinfo {year} {2016})}\BibitemShut {NoStop}%
\bibitem [{\citenamefont {Dominik}\ \emph {et~al.}(2012)\citenamefont
  {Dominik}, \citenamefont {Belczynski}, \citenamefont {Fryer}, \citenamefont
  {Holz}, \citenamefont {Berti}, \citenamefont {Bulik}, \citenamefont
  {Mandel},\ and\ \citenamefont {O'Shaughnessy}}]{Dominik:2012kk}%
  \BibitemOpen
  \bibfield  {author} {\bibinfo {author} {\bibfnamefont {M.}~\bibnamefont
  {Dominik}}, \bibinfo {author} {\bibfnamefont {K.}~\bibnamefont {Belczynski}},
  \bibinfo {author} {\bibfnamefont {C.}~\bibnamefont {Fryer}}, \bibinfo
  {author} {\bibfnamefont {D.}~\bibnamefont {Holz}}, \bibinfo {author}
  {\bibfnamefont {E.}~\bibnamefont {Berti}}, \bibinfo {author} {\bibfnamefont
  {T.}~\bibnamefont {Bulik}}, \bibinfo {author} {\bibfnamefont
  {I.}~\bibnamefont {Mandel}}, \ and\ \bibinfo {author} {\bibfnamefont
  {R.}~\bibnamefont {O'Shaughnessy}},\ }\href {\doibase
  10.1088/0004-637X/759/1/52} {\bibfield  {journal} {\bibinfo  {journal}
  {Astrophys. J.}\ }\textbf {\bibinfo {volume} {759}},\ \bibinfo {pages} {52}
  (\bibinfo {year} {2012})},\ \Eprint {http://arxiv.org/abs/1202.4901}
  {arXiv:1202.4901 [astro-ph.HE]} \BibitemShut {NoStop}%
\bibitem [{\citenamefont {Miller}(2016)}]{Miller:2016}%
  \BibitemOpen
  \bibfield  {author} {\bibinfo {author} {\bibfnamefont {M.~C.}\ \bibnamefont
  {Miller}},\ }\href {\doibase 10.3847/0004-637X/822/1/27} {\bibfield
  {journal} {\bibinfo  {journal} {Astrophys. J.}\ }\textbf {\bibinfo {volume}
  {822}},\ \bibinfo {pages} {27} (\bibinfo {year} {2016})}\BibitemShut
  {NoStop}%
\bibitem [{\citenamefont {Agathos}\ \emph {et~al.}(2015)\citenamefont
  {Agathos}, \citenamefont {Meidam}, \citenamefont {Del~Pozzo}, \citenamefont
  {Li}, \citenamefont {Tompitak}, \citenamefont {Veitch}, \citenamefont
  {Vitale},\ and\ \citenamefont {Broeck}}]{Agathos:2015uaa}%
  \BibitemOpen
  \bibfield  {author} {\bibinfo {author} {\bibfnamefont {M.}~\bibnamefont
  {Agathos}}, \bibinfo {author} {\bibfnamefont {J.}~\bibnamefont {Meidam}},
  \bibinfo {author} {\bibfnamefont {W.}~\bibnamefont {Del~Pozzo}}, \bibinfo
  {author} {\bibfnamefont {T.~G.~F.}\ \bibnamefont {Li}}, \bibinfo {author}
  {\bibfnamefont {M.}~\bibnamefont {Tompitak}}, \bibinfo {author}
  {\bibfnamefont {J.}~\bibnamefont {Veitch}}, \bibinfo {author} {\bibfnamefont
  {S.}~\bibnamefont {Vitale}}, \ and\ \bibinfo {author} {\bibfnamefont
  {C.~V.~D.}\ \bibnamefont {Broeck}},\ }\href {\doibase
  10.1103/PhysRevD.92.023012} {\bibfield  {journal} {\bibinfo  {journal} {Phys.
  Rev.}\ }\textbf {\bibinfo {volume} {D92}},\ \bibinfo {pages} {023012}
  (\bibinfo {year} {2015})},\ \Eprint {http://arxiv.org/abs/1503.05405}
  {arXiv:1503.05405 [gr-qc]} \BibitemShut {NoStop}%
\bibitem [{EPA()}]{EPAPS}%
  \BibitemOpen
  \href@noop {} {}\bibinfo {howpublished} {Supplementary Material}\BibitemShut
  {NoStop}%
\bibitem [{\citenamefont {Horowitz}\ \emph {et~al.}(2012)\citenamefont
  {Horowitz}, \citenamefont {Ahmed}, \citenamefont {Jen}, \citenamefont
  {Rakhman}, \citenamefont {Souder}, \citenamefont {Dalton}, \citenamefont
  {Liyanage}, \citenamefont {Paschke}, \citenamefont {Saenboonruang},
  \citenamefont {Silwal}, \citenamefont {Franklin}, \citenamefont {Friend},
  \citenamefont {Quinn}, \citenamefont {Kumar}, \citenamefont {McNulty},
  \citenamefont {Mercado}, \citenamefont {Riordan}, \citenamefont {Wexler},
  \citenamefont {Michaels},\ and\ \citenamefont {Urciuoli}}]{Horowitz:2012}%
  \BibitemOpen
  \bibfield  {author} {\bibinfo {author} {\bibfnamefont {C.~J.}\ \bibnamefont
  {Horowitz}}, \bibinfo {author} {\bibfnamefont {Z.}~\bibnamefont {Ahmed}},
  \bibinfo {author} {\bibfnamefont {C.-M.}\ \bibnamefont {Jen}}, \bibinfo
  {author} {\bibfnamefont {A.}~\bibnamefont {Rakhman}}, \bibinfo {author}
  {\bibfnamefont {P.~A.}\ \bibnamefont {Souder}}, \bibinfo {author}
  {\bibfnamefont {M.~M.}\ \bibnamefont {Dalton}}, \bibinfo {author}
  {\bibfnamefont {N.}~\bibnamefont {Liyanage}}, \bibinfo {author}
  {\bibfnamefont {K.~D.}\ \bibnamefont {Paschke}}, \bibinfo {author}
  {\bibfnamefont {K.}~\bibnamefont {Saenboonruang}}, \bibinfo {author}
  {\bibfnamefont {R.}~\bibnamefont {Silwal}}, \bibinfo {author} {\bibfnamefont
  {G.~B.}\ \bibnamefont {Franklin}}, \bibinfo {author} {\bibfnamefont
  {M.}~\bibnamefont {Friend}}, \bibinfo {author} {\bibfnamefont
  {B.}~\bibnamefont {Quinn}}, \bibinfo {author} {\bibfnamefont {K.~S.}\
  \bibnamefont {Kumar}}, \bibinfo {author} {\bibfnamefont {D.}~\bibnamefont
  {McNulty}}, \bibinfo {author} {\bibfnamefont {L.}~\bibnamefont {Mercado}},
  \bibinfo {author} {\bibfnamefont {S.}~\bibnamefont {Riordan}}, \bibinfo
  {author} {\bibfnamefont {J.}~\bibnamefont {Wexler}}, \bibinfo {author}
  {\bibfnamefont {R.~W.}\ \bibnamefont {Michaels}}, \ and\ \bibinfo {author}
  {\bibfnamefont {G.~M.}\ \bibnamefont {Urciuoli}},\ }\href {\doibase
  10.1103/PhysRevC.85.032501} {\bibfield  {journal} {\bibinfo  {journal} {Phys.
  Rev. C}\ }\textbf {\bibinfo {volume} {85}},\ \bibinfo {pages} {032501}
  (\bibinfo {year} {2012})}\BibitemShut {NoStop}%
\bibitem [{\citenamefont {Horowitz}\ \emph
  {et~al.}(2014{\natexlab{a}})\citenamefont {Horowitz}, \citenamefont {Kumar},\
  and\ \citenamefont {Michaels}}]{Horowitz:2014}%
  \BibitemOpen
  \bibfield  {author} {\bibinfo {author} {\bibfnamefont {C.}~\bibnamefont
  {Horowitz}}, \bibinfo {author} {\bibfnamefont {K.}~\bibnamefont {Kumar}}, \
  and\ \bibinfo {author} {\bibfnamefont {R.}~\bibnamefont {Michaels}},\ }\href
  {\doibase 10.1140/epja/i2014-14048-3} {\bibfield  {journal} {\bibinfo
  {journal} {Eur. Phys. J. A}\ }\textbf {\bibinfo {volume} {50}},\ \bibinfo
  {eid} {48} (\bibinfo {year} {2014}{\natexlab{a}}),\
  10.1140/epja/i2014-14048-3}\BibitemShut {NoStop}%
\bibitem [{\citenamefont {Horowitz}\ \emph
  {et~al.}(2014{\natexlab{b}})\citenamefont {Horowitz}, \citenamefont {Brown},
  \citenamefont {Kim}, \citenamefont {Lynch}, \citenamefont {Michaels},
  \citenamefont {Ono}, \citenamefont {Piekarewicz}, \citenamefont {Tsang},\
  and\ \citenamefont {Wolter}}]{Horowitz:2014a}%
  \BibitemOpen
  \bibfield  {author} {\bibinfo {author} {\bibfnamefont {C.~J.}\ \bibnamefont
  {Horowitz}}, \bibinfo {author} {\bibfnamefont {E.~F.}\ \bibnamefont {Brown}},
  \bibinfo {author} {\bibfnamefont {Y.}~\bibnamefont {Kim}}, \bibinfo {author}
  {\bibfnamefont {W.~G.}\ \bibnamefont {Lynch}}, \bibinfo {author}
  {\bibfnamefont {R.}~\bibnamefont {Michaels}}, \bibinfo {author}
  {\bibfnamefont {A.}~\bibnamefont {Ono}}, \bibinfo {author} {\bibfnamefont
  {J.}~\bibnamefont {Piekarewicz}}, \bibinfo {author} {\bibfnamefont {M.~B.}\
  \bibnamefont {Tsang}}, \ and\ \bibinfo {author} {\bibfnamefont {H.~H.}\
  \bibnamefont {Wolter}},\ }\href {\doibase 10.1088/0954-3899/41/9/093001}
  {\bibfield  {journal} {\bibinfo  {journal} {J. Phys. G}\ }\textbf {\bibinfo
  {volume} {41}},\ \bibinfo {pages} {093001} (\bibinfo {year}
  {2014}{\natexlab{b}})}\BibitemShut {NoStop}%
\bibitem [{\citenamefont {Hebeler}\ and\ \citenamefont
  {Schwenk}(2010)}]{Hebeler:2010}%
  \BibitemOpen
  \bibfield  {author} {\bibinfo {author} {\bibfnamefont {K.}~\bibnamefont
  {Hebeler}}\ and\ \bibinfo {author} {\bibfnamefont {A.}~\bibnamefont
  {Schwenk}},\ }\href {\doibase 10.1103/PhysRevC.82.014314} {\bibfield
  {journal} {\bibinfo  {journal} {Phys. Rev. C}\ }\textbf {\bibinfo {volume}
  {82}},\ \bibinfo {pages} {014314} (\bibinfo {year} {2010})}\BibitemShut
  {NoStop}%
\bibitem [{\citenamefont {Hebeler}\ \emph {et~al.}(2013)\citenamefont
  {Hebeler}, \citenamefont {Lattimer}, \citenamefont {Pethick},\ and\
  \citenamefont {Schwenk}}]{Hebeler:2013}%
  \BibitemOpen
  \bibfield  {author} {\bibinfo {author} {\bibfnamefont {K.}~\bibnamefont
  {Hebeler}}, \bibinfo {author} {\bibfnamefont {J.~M.}\ \bibnamefont
  {Lattimer}}, \bibinfo {author} {\bibfnamefont {C.~J.}\ \bibnamefont
  {Pethick}}, \ and\ \bibinfo {author} {\bibfnamefont {A.}~\bibnamefont
  {Schwenk}},\ }\href {\doibase 10.1088/0004-637X/773/1/11} {\bibfield
  {journal} {\bibinfo  {journal} {Astrophys. J.}\ }\textbf {\bibinfo {volume}
  {773}},\ \bibinfo {pages} {11} (\bibinfo {year} {2013})}\BibitemShut
  {NoStop}%
\bibitem [{\citenamefont {Wlaz{\l}owski}\ \emph {et~al.}(2014)\citenamefont
  {Wlaz{\l}owski}, \citenamefont {Holt}, \citenamefont {Moroz}, \citenamefont
  {Bulgac},\ and\ \citenamefont {Roche}}]{Wlazlowski:2014a}%
  \BibitemOpen
  \bibfield  {author} {\bibinfo {author} {\bibfnamefont {G.}~\bibnamefont
  {Wlaz{\l}owski}}, \bibinfo {author} {\bibfnamefont {J.~W.}\ \bibnamefont
  {Holt}}, \bibinfo {author} {\bibfnamefont {S.}~\bibnamefont {Moroz}},
  \bibinfo {author} {\bibfnamefont {A.}~\bibnamefont {Bulgac}}, \ and\ \bibinfo
  {author} {\bibfnamefont {K.~J.}\ \bibnamefont {Roche}},\ }\href {\doibase
  10.1103/PhysRevLett.113.182503} {\bibfield  {journal} {\bibinfo  {journal}
  {Phys. Rev. Lett.}\ }\textbf {\bibinfo {volume} {113}},\ \bibinfo {pages}
  {182503} (\bibinfo {year} {2014})},\ \Eprint {http://arxiv.org/abs/1403.3753}
  {arXiv:1403.3753} \BibitemShut {NoStop}%
\bibitem [{\citenamefont {Gandolfi}\ \emph
  {et~al.}(2014{\natexlab{b}})\citenamefont {Gandolfi}, \citenamefont {Lovato},
  \citenamefont {Carlson},\ and\ \citenamefont {Schmidt}}]{Gandolfi:2014}%
  \BibitemOpen
  \bibfield  {author} {\bibinfo {author} {\bibfnamefont {S.}~\bibnamefont
  {Gandolfi}}, \bibinfo {author} {\bibfnamefont {A.}~\bibnamefont {Lovato}},
  \bibinfo {author} {\bibfnamefont {J.}~\bibnamefont {Carlson}}, \ and\
  \bibinfo {author} {\bibfnamefont {K.~E.}\ \bibnamefont {Schmidt}},\ }\href
  {\doibase 10.1103/PhysRevC.90.061306} {\bibfield  {journal} {\bibinfo
  {journal} {Phys. Rev. C}\ }\textbf {\bibinfo {volume} {90}},\ \bibinfo
  {pages} {061306} (\bibinfo {year} {2014}{\natexlab{b}})},\ \Eprint
  {http://arxiv.org/abs/1406.3388} {arXiv:1406.3388 [nucl-th]} \BibitemShut
  {NoStop}%
\bibitem [{\citenamefont {Lynn}\ \emph {et~al.}(2015)\citenamefont {Lynn},
  \citenamefont {Tews}, \citenamefont {Carlson}, \citenamefont {Gandolfi},
  \citenamefont {Gezerlis}, \citenamefont {Schmidt},\ and\ \citenamefont
  {Schwenk}}]{Lynn:2015}%
  \BibitemOpen
  \bibfield  {author} {\bibinfo {author} {\bibfnamefont {J.~E.}\ \bibnamefont
  {Lynn}}, \bibinfo {author} {\bibfnamefont {I.}~\bibnamefont {Tews}}, \bibinfo
  {author} {\bibfnamefont {J.}~\bibnamefont {Carlson}}, \bibinfo {author}
  {\bibfnamefont {S.}~\bibnamefont {Gandolfi}}, \bibinfo {author}
  {\bibfnamefont {A.}~\bibnamefont {Gezerlis}}, \bibinfo {author}
  {\bibfnamefont {K.~E.}\ \bibnamefont {Schmidt}}, \ and\ \bibinfo {author}
  {\bibfnamefont {A.}~\bibnamefont {Schwenk}},\ }\href
  {http://arxiv.org/abs/1509.03470} {\enquote {\bibinfo {title} {Chiral
  three-nucleon interactions in light nuclei, neutron-{$\alpha$} scattering,
  and neutron matter},}\ } (\bibinfo {year} {2015}),\ \Eprint
  {http://arxiv.org/abs/arXiv:1509.03470} {arXiv:1509.03470} \BibitemShut
  {NoStop}%
\bibitem [{\citenamefont {Page}\ and\ \citenamefont {Reddy}(2006)}]{Page:2006}%
  \BibitemOpen
  \bibfield  {author} {\bibinfo {author} {\bibfnamefont {D.}~\bibnamefont
  {Page}}\ and\ \bibinfo {author} {\bibfnamefont {S.}~\bibnamefont {Reddy}},\
  }\href {\doibase 10.1146/annurev.nucl.56.080805.140600} {\bibfield  {journal}
  {\bibinfo  {journal} {Annu. Rev. Nucl. Part. Sci.}\ }\textbf {\bibinfo
  {volume} {56}},\ \bibinfo {pages} {327} (\bibinfo {year} {2006})},\ \Eprint
  {http://arxiv.org/abs/astro-ph/0608360v1} {arXiv:astro-ph/0608360v1}
  \BibitemShut {NoStop}%
\bibitem [{\citenamefont {Steiner}\ and\ \citenamefont
  {Gandolfi}(2012)}]{Steiner:2012}%
  \BibitemOpen
  \bibfield  {author} {\bibinfo {author} {\bibfnamefont {A.~W.}\ \bibnamefont
  {Steiner}}\ and\ \bibinfo {author} {\bibfnamefont {S.}~\bibnamefont
  {Gandolfi}},\ }\href {\doibase 10.1103/PhysRevLett.108.081102} {\bibfield
  {journal} {\bibinfo  {journal} {Phys. Rev. Lett.}\ }\textbf {\bibinfo
  {volume} {108}},\ \bibinfo {pages} {081102} (\bibinfo {year}
  {2012})}\BibitemShut {NoStop}%
\bibitem [{\citenamefont {Steiner}\ \emph {et~al.}(2013)\citenamefont
  {Steiner}, \citenamefont {Lattimer},\ and\ \citenamefont
  {Brown}}]{Steiner:2013}%
  \BibitemOpen
  \bibfield  {author} {\bibinfo {author} {\bibfnamefont {A.~W.}\ \bibnamefont
  {Steiner}}, \bibinfo {author} {\bibfnamefont {J.~M.}\ \bibnamefont
  {Lattimer}}, \ and\ \bibinfo {author} {\bibfnamefont {E.~F.}\ \bibnamefont
  {Brown}},\ }\href {\doibase 10.1088/2041-8205/765/1/L5} {\bibfield  {journal}
  {\bibinfo  {journal} {Astrophys. J. Lett.}\ }\textbf {\bibinfo {volume}
  {765}},\ \bibinfo {pages} {L5} (\bibinfo {year} {2013})}\BibitemShut
  {NoStop}%
\bibitem [{\citenamefont {Lattimer}\ and\ \citenamefont
  {Steiner}(2014)}]{Lattimer:2014}%
  \BibitemOpen
  \bibfield  {author} {\bibinfo {author} {\bibfnamefont {J.~M.}\ \bibnamefont
  {Lattimer}}\ and\ \bibinfo {author} {\bibfnamefont {A.~W.}\ \bibnamefont
  {Steiner}},\ }\href {\doibase 10.1140/epja/i2014-14040-y} {\bibfield
  {journal} {\bibinfo  {journal} {Eur. Phys. J. A}\ }\textbf {\bibinfo {volume}
  {50}},\ \bibinfo {pages} {1} (\bibinfo {year} {2014})}\BibitemShut {NoStop}%
\bibitem [{\citenamefont {{Baym}}\ \emph {et~al.}(1971)\citenamefont {{Baym}},
  \citenamefont {{Pethick}},\ and\ \citenamefont {{Sutherland}}}]{Baym:1971}%
  \BibitemOpen
  \bibfield  {author} {\bibinfo {author} {\bibfnamefont {G.}~\bibnamefont
  {{Baym}}}, \bibinfo {author} {\bibfnamefont {C.}~\bibnamefont {{Pethick}}}, \
  and\ \bibinfo {author} {\bibfnamefont {P.}~\bibnamefont {{Sutherland}}},\
  }\href {\doibase 10.1086/151216} {\bibfield  {journal} {\bibinfo  {journal}
  {Astrophys. J.}\ }\textbf {\bibinfo {volume} {170}},\ \bibinfo {pages} {299}
  (\bibinfo {year} {1971})}\BibitemShut {NoStop}%
\bibitem [{\citenamefont {Negele}\ and\ \citenamefont
  {Vautherin}(1973)}]{Negele:1973}%
  \BibitemOpen
  \bibfield  {author} {\bibinfo {author} {\bibfnamefont {J.~W.}\ \bibnamefont
  {Negele}}\ and\ \bibinfo {author} {\bibfnamefont {D.}~\bibnamefont
  {Vautherin}},\ }\href {\doibase 10.1016/0375-9474(73)90349-7} {\bibfield
  {journal} {\bibinfo  {journal} {Nucl. Phys. A}\ }\textbf {\bibinfo {volume}
  {207}},\ \bibinfo {pages} {298} (\bibinfo {year} {1973})}\BibitemShut
  {NoStop}%
\bibitem [{\citenamefont {Sharma}\ \emph {et~al.}(2015)\citenamefont {Sharma},
  \citenamefont {Centelles}, \citenamefont {Vinas}, \citenamefont {Baldo},\
  and\ \citenamefont {Burgio}}]{Sharma:2015}%
  \BibitemOpen
  \bibfield  {author} {\bibinfo {author} {\bibfnamefont {B.~K.}\ \bibnamefont
  {Sharma}}, \bibinfo {author} {\bibfnamefont {M.}~\bibnamefont {Centelles}},
  \bibinfo {author} {\bibfnamefont {X.}~\bibnamefont {Vinas}}, \bibinfo
  {author} {\bibfnamefont {M.}~\bibnamefont {Baldo}}, \ and\ \bibinfo {author}
  {\bibfnamefont {G.~F.}\ \bibnamefont {Burgio}},\ }\href {\doibase
  10.1051/0004-6361/201526642} {\bibfield  {journal} {\bibinfo  {journal}
  {Astron. \& Astrophys.}\ }\textbf {\bibinfo {volume} {584}},\ \bibinfo
  {pages} {A103} (\bibinfo {year} {2015})},\ \Eprint
  {http://arxiv.org/abs/1506.00375} {arXiv:1506.00375} \BibitemShut {NoStop}%
\bibitem [{\citenamefont {Haensel}\ \emph {et~al.}(2007)\citenamefont
  {Haensel}, \citenamefont {Potekhin},\ and\ \citenamefont
  {Yakovlev}}]{Haensel:2007}%
  \BibitemOpen
  \bibfield  {author} {\bibinfo {author} {\bibfnamefont {P.}~\bibnamefont
  {Haensel}}, \bibinfo {author} {\bibfnamefont {A.~Y.}\ \bibnamefont
  {Potekhin}}, \ and\ \bibinfo {author} {\bibfnamefont {D.~G.}\ \bibnamefont
  {Yakovlev}},\ }\href {\doibase 10.1007/978-0-387-47301-7} {\emph {\bibinfo
  {title} {Neutron Stars 1}}},\ \bibinfo {edition} {1st}\ ed.,\ \bibinfo
  {series} {Astrophysics and Space Science Library}, Vol.\ \bibinfo {volume}
  {326}\ (\bibinfo  {publisher} {Springer-Verlag},\ \bibinfo {address} {New
  York},\ \bibinfo {year} {2007})\BibitemShut {NoStop}%
\bibitem [{\citenamefont {Chamel}\ and\ \citenamefont
  {Haensel}(2008)}]{Chamel:2008}%
  \BibitemOpen
  \bibfield  {author} {\bibinfo {author} {\bibfnamefont {N.}~\bibnamefont
  {Chamel}}\ and\ \bibinfo {author} {\bibfnamefont {P.}~\bibnamefont
  {Haensel}},\ }\href {\doibase 10.12942/lrr-2008-10} {\bibfield  {journal}
  {\bibinfo  {journal} {Living Rev. Relativity}\ }\textbf {\bibinfo {volume}
  {11}} (\bibinfo {year} {2008}),\ 10.12942/lrr-2008-10},\ \Eprint
  {http://arxiv.org/abs/0812.3955} {arXiv:0812.3955} \BibitemShut {NoStop}%
\bibitem [{\citenamefont {Nelson}\ \emph {et~al.}(2018)\citenamefont {Nelson},
  \citenamefont {Reddy},\ and\ \citenamefont {Zhou}}]{Nelson:2018xtr}%
  \BibitemOpen
  \bibfield  {author} {\bibinfo {author} {\bibfnamefont {A.}~\bibnamefont
  {Nelson}}, \bibinfo {author} {\bibfnamefont {S.}~\bibnamefont {Reddy}}, \
  and\ \bibinfo {author} {\bibfnamefont {D.}~\bibnamefont {Zhou}},\ }\href@noop
  {} {\  (\bibinfo {year} {2018})},\ \Eprint {http://arxiv.org/abs/1803.03266}
  {arXiv:1803.03266 [hep-ph]} \BibitemShut {NoStop}%
\bibitem [{\citenamefont {Read}\ \emph {et~al.}(2009)\citenamefont {Read},
  \citenamefont {Lackey}, \citenamefont {Owen},\ and\ \citenamefont
  {Friedman}}]{Read:2009}%
  \BibitemOpen
  \bibfield  {author} {\bibinfo {author} {\bibfnamefont {J.~S.}\ \bibnamefont
  {Read}}, \bibinfo {author} {\bibfnamefont {B.~D.}\ \bibnamefont {Lackey}},
  \bibinfo {author} {\bibfnamefont {B.~J.}\ \bibnamefont {Owen}}, \ and\
  \bibinfo {author} {\bibfnamefont {J.~L.}\ \bibnamefont {Friedman}},\ }\href
  {\doibase 10.1103/PhysRevD.79.124032} {\bibfield  {journal} {\bibinfo
  {journal} {Phys. Rev. D}\ }\textbf {\bibinfo {volume} {79}},\ \bibinfo
  {pages} {124032} (\bibinfo {year} {2009})}\BibitemShut {NoStop}%
\bibitem [{\citenamefont {Fortin}\ \emph {et~al.}(2016)\citenamefont {Fortin},
  \citenamefont {Provid{\^e}ncia}, \citenamefont {Raduta}, \citenamefont
  {Gulminelli}, \citenamefont {Zdunik}, \citenamefont {Haensel},\ and\
  \citenamefont {Bejger}}]{Fortin:2016}%
  \BibitemOpen
  \bibfield  {author} {\bibinfo {author} {\bibfnamefont {M.}~\bibnamefont
  {Fortin}}, \bibinfo {author} {\bibfnamefont {C.}~\bibnamefont
  {Provid{\^e}ncia}}, \bibinfo {author} {\bibfnamefont {A.~R.}\ \bibnamefont
  {Raduta}}, \bibinfo {author} {\bibfnamefont {F.}~\bibnamefont {Gulminelli}},
  \bibinfo {author} {\bibfnamefont {J.~L.}\ \bibnamefont {Zdunik}}, \bibinfo
  {author} {\bibfnamefont {P.}~\bibnamefont {Haensel}}, \ and\ \bibinfo
  {author} {\bibfnamefont {M.}~\bibnamefont {Bejger}},\ }\href {\doibase
  10.1103/PhysRevC.94.035804} {\bibfield  {journal} {\bibinfo  {journal} {Phys.
  Rev. C}\ }\textbf {\bibinfo {volume} {94}},\ \bibinfo {pages} {035804}
  (\bibinfo {year} {2016})},\ \Eprint {http://arxiv.org/abs/1604.01944}
  {arXiv:1604.01944} \BibitemShut {NoStop}%
\bibitem [{\citenamefont {Zdunik}\ \emph {et~al.}(2016)\citenamefont {Zdunik},
  \citenamefont {Fortin},\ and\ \citenamefont {Haensel}}]{Zdunik:2016}%
  \BibitemOpen
  \bibfield  {author} {\bibinfo {author} {\bibfnamefont {J.~L.}\ \bibnamefont
  {Zdunik}}, \bibinfo {author} {\bibfnamefont {M.}~\bibnamefont {Fortin}}, \
  and\ \bibinfo {author} {\bibfnamefont {P.}~\bibnamefont {Haensel}},\
  }\href@noop {} {\enquote {\bibinfo {title} {Neutron star properties and the
  equation of state for its core},}\ } (\bibinfo {year} {2016}),\ \Eprint
  {http://arxiv.org/abs/1611.01357} {arXiv:1611.01357} \BibitemShut {NoStop}%
\bibitem [{\citenamefont {Ravenhall}\ \emph {et~al.}(1983)\citenamefont
  {Ravenhall}, \citenamefont {Pethick},\ and\ \citenamefont
  {Wilson}}]{Ravenhall:1983uh}%
  \BibitemOpen
  \bibfield  {author} {\bibinfo {author} {\bibfnamefont {D.~G.}\ \bibnamefont
  {Ravenhall}}, \bibinfo {author} {\bibfnamefont {C.~J.}\ \bibnamefont
  {Pethick}}, \ and\ \bibinfo {author} {\bibfnamefont {J.~R.}\ \bibnamefont
  {Wilson}},\ }\href {\doibase 10.1103/PhysRevLett.50.2066} {\bibfield
  {journal} {\bibinfo  {journal} {Phys. Rev. Lett.}\ }\textbf {\bibinfo
  {volume} {50}},\ \bibinfo {pages} {2066} (\bibinfo {year}
  {1983})}\BibitemShut {NoStop}%
\bibitem [{\citenamefont {Chamel}\ \emph {et~al.}(2007)\citenamefont {Chamel},
  \citenamefont {Naimi}, \citenamefont {Khan},\ and\ \citenamefont
  {Margueron}}]{Chamel:2007}%
  \BibitemOpen
  \bibfield  {author} {\bibinfo {author} {\bibfnamefont {N.}~\bibnamefont
  {Chamel}}, \bibinfo {author} {\bibfnamefont {S.}~\bibnamefont {Naimi}},
  \bibinfo {author} {\bibfnamefont {E.}~\bibnamefont {Khan}}, \ and\ \bibinfo
  {author} {\bibfnamefont {J.}~\bibnamefont {Margueron}},\ }\href {\doibase
  10.1103/PhysRevC.75.055806} {\bibfield  {journal} {\bibinfo  {journal} {Phys.
  Rev. C}\ }\textbf {\bibinfo {volume} {75}},\ \bibinfo {pages} {055806}
  (\bibinfo {year} {2007})}\BibitemShut {NoStop}%
\bibitem [{\citenamefont {Lattimer}\ \emph {et~al.}(1985)\citenamefont
  {Lattimer}, \citenamefont {Pethick}, \citenamefont {Ravenhall},\ and\
  \citenamefont {Lamb}}]{Lattimer:1985}%
  \BibitemOpen
  \bibfield  {author} {\bibinfo {author} {\bibfnamefont {J.}~\bibnamefont
  {Lattimer}}, \bibinfo {author} {\bibfnamefont {C.}~\bibnamefont {Pethick}},
  \bibinfo {author} {\bibfnamefont {D.}~\bibnamefont {Ravenhall}}, \ and\
  \bibinfo {author} {\bibfnamefont {D.}~\bibnamefont {Lamb}},\ }\href {\doibase
  10.1016/0375-9474(85)90006-5} {\bibfield  {journal} {\bibinfo  {journal}
  {Nucl. Phys. A}\ }\textbf {\bibinfo {volume} {432}},\ \bibinfo {pages} {646 }
  (\bibinfo {year} {1985})}\BibitemShut {NoStop}%
\bibitem [{\citenamefont {Steiner}(2012)}]{Steiner:2012a}%
  \BibitemOpen
  \bibfield  {author} {\bibinfo {author} {\bibfnamefont {A.~W.}\ \bibnamefont
  {Steiner}},\ }\href {\doibase 10.1103/PhysRevC.85.055804} {\bibfield
  {journal} {\bibinfo  {journal} {Phys. Rev. C}\ }\textbf {\bibinfo {volume}
  {85}},\ \bibinfo {pages} {055804} (\bibinfo {year} {2012})}\BibitemShut
  {NoStop}%
\bibitem [{\citenamefont {Margueron}\ \emph {et~al.}(2018)\citenamefont
  {Margueron}, \citenamefont {Hoffmann~Casali},\ and\ \citenamefont
  {Gulminelli}}]{Margueron:2018}%
  \BibitemOpen
  \bibfield  {author} {\bibinfo {author} {\bibfnamefont {J.}~\bibnamefont
  {Margueron}}, \bibinfo {author} {\bibfnamefont {R.}~\bibnamefont
  {Hoffmann~Casali}}, \ and\ \bibinfo {author} {\bibfnamefont {F.}~\bibnamefont
  {Gulminelli}},\ }\href {\doibase 10.1103/PhysRevC.97.025805} {\bibfield
  {journal} {\bibinfo  {journal} {Phys. Rev. C}\ }\textbf {\bibinfo {volume}
  {97}},\ \bibinfo {pages} {025805} (\bibinfo {year} {2018})}\BibitemShut
  {NoStop}%
\bibitem [{\citenamefont {Gandolfi}\ \emph {et~al.}(2015)\citenamefont
  {Gandolfi}, \citenamefont {Gezerlis},\ and\ \citenamefont
  {Carlson}}]{Gandolfi:2015}%
  \BibitemOpen
  \bibfield  {author} {\bibinfo {author} {\bibfnamefont {S.}~\bibnamefont
  {Gandolfi}}, \bibinfo {author} {\bibfnamefont {A.}~\bibnamefont {Gezerlis}},
  \ and\ \bibinfo {author} {\bibfnamefont {J.}~\bibnamefont {Carlson}},\ }\href
  {\doibase 10.1146/annurev-nucl-102014-021957} {\bibfield  {journal} {\bibinfo
   {journal} {Annu. Rev. Nucl. Part. Sci.}\ }\textbf {\bibinfo {volume} {65}},\
  \bibinfo {pages} {303} (\bibinfo {year} {2015})},\ \Eprint
  {http://arxiv.org/abs/1501.0567} {arXiv:1501.0567} \BibitemShut {NoStop}%
\bibitem [{\citenamefont {Roggero}\ \emph {et~al.}(2014)\citenamefont
  {Roggero}, \citenamefont {Mukherjee},\ and\ \citenamefont
  {Pederiva}}]{Roggero:2014}%
  \BibitemOpen
  \bibfield  {author} {\bibinfo {author} {\bibfnamefont {A.}~\bibnamefont
  {Roggero}}, \bibinfo {author} {\bibfnamefont {A.}~\bibnamefont {Mukherjee}},
  \ and\ \bibinfo {author} {\bibfnamefont {F.}~\bibnamefont {Pederiva}},\
  }\href {\doibase 10.1103/PhysRevLett.112.221103} {\bibfield  {journal}
  {\bibinfo  {journal} {Phys. Rev. Lett.}\ }\textbf {\bibinfo {volume} {112}},\
  \bibinfo {pages} {221103} (\bibinfo {year} {2014})}\BibitemShut {NoStop}%
\bibitem [{\citenamefont {Rrapaj}\ \emph {et~al.}(2016)\citenamefont {Rrapaj},
  \citenamefont {Roggero},\ and\ \citenamefont {Holt}}]{Rrapaj:2016}%
  \BibitemOpen
  \bibfield  {author} {\bibinfo {author} {\bibfnamefont {E.}~\bibnamefont
  {Rrapaj}}, \bibinfo {author} {\bibfnamefont {A.}~\bibnamefont {Roggero}}, \
  and\ \bibinfo {author} {\bibfnamefont {J.~W.}\ \bibnamefont {Holt}},\ }\href
  {\doibase 10.1103/PhysRevC.93.065801} {\bibfield  {journal} {\bibinfo
  {journal} {prc}\ }\textbf {\bibinfo {volume} {93}},\ \bibinfo {pages}
  {065801} (\bibinfo {year} {2016})}\BibitemShut {NoStop}%
\bibitem [{\citenamefont {Lee}\ \emph {et~al.}(1998)\citenamefont {Lee},
  \citenamefont {Kuo}, \citenamefont {Li},\ and\ \citenamefont
  {Brown}}]{Lee:1998}%
  \BibitemOpen
  \bibfield  {author} {\bibinfo {author} {\bibfnamefont {C.-H.}\ \bibnamefont
  {Lee}}, \bibinfo {author} {\bibfnamefont {T.~T.~S.}\ \bibnamefont {Kuo}},
  \bibinfo {author} {\bibfnamefont {G.~Q.}\ \bibnamefont {Li}}, \ and\ \bibinfo
  {author} {\bibfnamefont {G.~E.}\ \bibnamefont {Brown}},\ }\href {\doibase
  10.1103/PhysRevC.57.3488} {\bibfield  {journal} {\bibinfo  {journal} {Phys.
  Rev. C}\ }\textbf {\bibinfo {volume} {57}},\ \bibinfo {pages} {3488}
  (\bibinfo {year} {1998})}\BibitemShut {NoStop}%
\bibitem [{\citenamefont {Gonzalez-Boquera}\ \emph {et~al.}(2017)\citenamefont
  {Gonzalez-Boquera}, \citenamefont {Centelles}, \citenamefont {Vi\~nas},\ and\
  \citenamefont {Rios}}]{Gonzalez-Boquera:2017}%
  \BibitemOpen
  \bibfield  {author} {\bibinfo {author} {\bibfnamefont {C.}~\bibnamefont
  {Gonzalez-Boquera}}, \bibinfo {author} {\bibfnamefont {M.}~\bibnamefont
  {Centelles}}, \bibinfo {author} {\bibfnamefont {X.}~\bibnamefont {Vi\~nas}},
  \ and\ \bibinfo {author} {\bibfnamefont {A.}~\bibnamefont {Rios}},\ }\href
  {\doibase 10.1103/PhysRevC.96.065806} {\bibfield  {journal} {\bibinfo
  {journal} {Phys. Rev. C}\ }\textbf {\bibinfo {volume} {96}},\ \bibinfo
  {pages} {065806} (\bibinfo {year} {2017})}\BibitemShut {NoStop}%
\bibitem [{\citenamefont {Li}\ \emph {et~al.}(2008)\citenamefont {Li},
  \citenamefont {Chen},\ and\ \citenamefont {Ko}}]{Li:2008}%
  \BibitemOpen
  \bibfield  {author} {\bibinfo {author} {\bibfnamefont {B.-A.}\ \bibnamefont
  {Li}}, \bibinfo {author} {\bibfnamefont {L.-W.}\ \bibnamefont {Chen}}, \ and\
  \bibinfo {author} {\bibfnamefont {C.~M.}\ \bibnamefont {Ko}},\ }\href
  {\doibase https://doi.org/10.1016/j.physrep.2008.04.005} {\bibfield
  {journal} {\bibinfo  {journal} {Phys. Rep.}\ }\textbf {\bibinfo {volume}
  {464}},\ \bibinfo {pages} {113 } (\bibinfo {year} {2008})}\BibitemShut
  {NoStop}%
\bibitem [{\citenamefont {Bulgac}\ \emph {et~al.}(2018)\citenamefont {Bulgac},
  \citenamefont {Forbes}, \citenamefont {Jin}, \citenamefont {Perez},\ and\
  \citenamefont {Schunck}}]{Bulgac:2018}%
  \BibitemOpen
  \bibfield  {author} {\bibinfo {author} {\bibfnamefont {A.}~\bibnamefont
  {Bulgac}}, \bibinfo {author} {\bibfnamefont {M.~M.}\ \bibnamefont {Forbes}},
  \bibinfo {author} {\bibfnamefont {S.}~\bibnamefont {Jin}}, \bibinfo {author}
  {\bibfnamefont {R.~N.}\ \bibnamefont {Perez}}, \ and\ \bibinfo {author}
  {\bibfnamefont {N.}~\bibnamefont {Schunck}},\ }\href {\doibase
  10.1103/PhysRevC.97.044313} {\bibfield  {journal} {\bibinfo  {journal} {Phys.
  Rev. C}\ }\textbf {\bibinfo {volume} {97}},\ \bibinfo {pages} {044313}
  (\bibinfo {year} {2018})},\ \Eprint {http://arxiv.org/abs/1708.08771}
  {arXiv:1708.08771 [nucl-th]} \BibitemShut {NoStop}%
\bibitem [{\citenamefont {Alford}()}]{Alford:2016pc}%
  \BibitemOpen
  \bibfield  {author} {\bibinfo {author} {\bibfnamefont {M.}~\bibnamefont
  {Alford}},\ }\href {http://www.int.washington.edu/PROGRAMS/16-2b/} {}\bibinfo
  {note} {Raised in discussions at the
  \href{http://www.int.washington.edu/PROGRAMS/16-2b/}{INT-16-2b}
  program.}\BibitemShut {Stop}%
\bibitem [{\citenamefont {Vines}\ \emph {et~al.}(2011)\citenamefont {Vines},
  \citenamefont {Flanagan},\ and\ \citenamefont {Hinderer}}]{Vines:2011ud}%
  \BibitemOpen
  \bibfield  {author} {\bibinfo {author} {\bibfnamefont {J.}~\bibnamefont
  {Vines}}, \bibinfo {author} {\bibfnamefont {E.~E.}\ \bibnamefont {Flanagan}},
  \ and\ \bibinfo {author} {\bibfnamefont {T.}~\bibnamefont {Hinderer}},\
  }\href {\doibase 10.1103/PhysRevD.83.084051} {\bibfield  {journal} {\bibinfo
  {journal} {Phys. Rev.}\ }\textbf {\bibinfo {volume} {D83}},\ \bibinfo {pages}
  {084051} (\bibinfo {year} {2011})},\ \Eprint {http://arxiv.org/abs/1101.1673}
  {arXiv:1101.1673 [gr-qc]} \BibitemShut {NoStop}%
\bibitem [{\citenamefont {{Del Pozzo}}\ \emph {et~al.}(2013)\citenamefont {{Del
  Pozzo}}, \citenamefont {{Li}}, \citenamefont {{Agathos}}, \citenamefont {{Van
  Den Broeck}},\ and\ \citenamefont {{Vitale}}}]{DelPozzo:2013PRL}%
  \BibitemOpen
  \bibfield  {author} {\bibinfo {author} {\bibfnamefont {W.}~\bibnamefont {{Del
  Pozzo}}}, \bibinfo {author} {\bibfnamefont {T.~G.~F.}\ \bibnamefont {{Li}}},
  \bibinfo {author} {\bibfnamefont {M.}~\bibnamefont {{Agathos}}}, \bibinfo
  {author} {\bibfnamefont {C.}~\bibnamefont {{Van Den Broeck}}}, \ and\
  \bibinfo {author} {\bibfnamefont {S.}~\bibnamefont {{Vitale}}},\ }\href
  {\doibase 10.1103/PhysRevLett.111.071101} {\bibfield  {journal} {\bibinfo
  {journal} {Physical Review Letters}\ }\textbf {\bibinfo {volume} {111}},\
  \bibinfo {eid} {071101} (\bibinfo {year} {2013})},\ \Eprint
  {http://arxiv.org/abs/1307.8338} {arXiv:1307.8338 [gr-qc]} \BibitemShut
  {NoStop}%
\bibitem [{\citenamefont {Bose}\ \emph {et~al.}(2018)\citenamefont {Bose},
  \citenamefont {Chakravarti}, \citenamefont {Rezzolla}, \citenamefont
  {Sathyaprakash},\ and\ \citenamefont {Takami}}]{Bose:2017jvk}%
  \BibitemOpen
  \bibfield  {author} {\bibinfo {author} {\bibfnamefont {S.}~\bibnamefont
  {Bose}}, \bibinfo {author} {\bibfnamefont {K.}~\bibnamefont {Chakravarti}},
  \bibinfo {author} {\bibfnamefont {L.}~\bibnamefont {Rezzolla}}, \bibinfo
  {author} {\bibfnamefont {B.~S.}\ \bibnamefont {Sathyaprakash}}, \ and\
  \bibinfo {author} {\bibfnamefont {K.}~\bibnamefont {Takami}},\ }\href
  {\doibase 10.1103/PhysRevLett.120.031102} {\bibfield  {journal} {\bibinfo
  {journal} {Phys. Rev. Lett.}\ }\textbf {\bibinfo {volume} {120}},\ \bibinfo
  {pages} {031102} (\bibinfo {year} {2018})},\ \Eprint
  {http://arxiv.org/abs/1705.10850} {arXiv:1705.10850 [gr-qc]} \BibitemShut
  {NoStop}%
\bibitem [{\citenamefont {Abbott}\ \emph
  {et~al.}(2018{\natexlab{b}})\citenamefont {Abbott} \emph
  {et~al.}}]{LIGOScientific:2018mvr}%
  \BibitemOpen
  \bibfield  {author} {\bibinfo {author} {\bibfnamefont {B.~P.}\ \bibnamefont
  {Abbott}} \emph {et~al.} (\bibinfo {collaboration} {LIGO Scientific,
  Virgo}),\ }\href@noop {} {\  (\bibinfo {year} {2018}{\natexlab{b}})},\
  \Eprint {http://arxiv.org/abs/1811.12907} {arXiv:1811.12907 [astro-ph.HE]}
  \BibitemShut {NoStop}%
\bibitem [{\citenamefont {Stovall}\ \emph {et~al.}(2018)\citenamefont {Stovall}
  \emph {et~al.}}]{Stovall:2018ouw}%
  \BibitemOpen
  \bibfield  {author} {\bibinfo {author} {\bibfnamefont {K.}~\bibnamefont
  {Stovall}} \emph {et~al.},\ }\href {\doibase 10.3847/2041-8213/aaad06}
  {\bibfield  {journal} {\bibinfo  {journal} {Astrophys. J.}\ }\textbf
  {\bibinfo {volume} {854}},\ \bibinfo {pages} {L22} (\bibinfo {year}
  {2018})},\ \Eprint {http://arxiv.org/abs/1802.01707} {arXiv:1802.01707
  [astro-ph.HE]} \BibitemShut {NoStop}%
\bibitem [{\citenamefont {Buonanno}\ \emph {et~al.}(2009)\citenamefont
  {Buonanno}, \citenamefont {Iyer}, \citenamefont {Ochsner}, \citenamefont
  {Pan},\ and\ \citenamefont {Sathyaprakash}}]{Buonanno:2009zt}%
  \BibitemOpen
  \bibfield  {author} {\bibinfo {author} {\bibfnamefont {A.}~\bibnamefont
  {Buonanno}}, \bibinfo {author} {\bibfnamefont {B.}~\bibnamefont {Iyer}},
  \bibinfo {author} {\bibfnamefont {E.}~\bibnamefont {Ochsner}}, \bibinfo
  {author} {\bibfnamefont {Y.}~\bibnamefont {Pan}}, \ and\ \bibinfo {author}
  {\bibfnamefont {B.~S.}\ \bibnamefont {Sathyaprakash}},\ }\href {\doibase
  10.1103/PhysRevD.80.084043} {\bibfield  {journal} {\bibinfo  {journal} {Phys.
  Rev.}\ }\textbf {\bibinfo {volume} {D80}},\ \bibinfo {pages} {084043}
  (\bibinfo {year} {2009})},\ \Eprint {http://arxiv.org/abs/0907.0700}
  {arXiv:0907.0700 [gr-qc]} \BibitemShut {NoStop}%
\bibitem [{\citenamefont {Dietrich}\ \emph {et~al.}(2017)\citenamefont
  {Dietrich}, \citenamefont {Bernuzzi},\ and\ \citenamefont
  {Tichy}}]{Dietrich:2017aum}%
  \BibitemOpen
  \bibfield  {author} {\bibinfo {author} {\bibfnamefont {T.}~\bibnamefont
  {Dietrich}}, \bibinfo {author} {\bibfnamefont {S.}~\bibnamefont {Bernuzzi}},
  \ and\ \bibinfo {author} {\bibfnamefont {W.}~\bibnamefont {Tichy}},\ }\href
  {\doibase 10.1103/PhysRevD.96.121501} {\bibfield  {journal} {\bibinfo
  {journal} {Phys. Rev.}\ }\textbf {\bibinfo {volume} {D96}},\ \bibinfo {pages}
  {121501} (\bibinfo {year} {2017})},\ \Eprint
  {http://arxiv.org/abs/1706.02969} {arXiv:1706.02969 [gr-qc]} \BibitemShut
  {NoStop}%
\bibitem [{\citenamefont {Hinderer}(2008)}]{Hinderer:2007mb}%
  \BibitemOpen
  \bibfield  {author} {\bibinfo {author} {\bibfnamefont {T.}~\bibnamefont
  {Hinderer}},\ }\href {\doibase 10.1086/533487} {\bibfield  {journal}
  {\bibinfo  {journal} {Astrophys. J.}\ }\textbf {\bibinfo {volume} {677}},\
  \bibinfo {pages} {1216} (\bibinfo {year} {2008})},\ \Eprint
  {http://arxiv.org/abs/0711.2420} {arXiv:0711.2420 [astro-ph]} \BibitemShut
  {NoStop}%
\bibitem [{\citenamefont {Postnikov}\ \emph {et~al.}(2010)\citenamefont
  {Postnikov}, \citenamefont {Prakash},\ and\ \citenamefont
  {Lattimer}}]{Postnikov:2010}%
  \BibitemOpen
  \bibfield  {author} {\bibinfo {author} {\bibfnamefont {S.}~\bibnamefont
  {Postnikov}}, \bibinfo {author} {\bibfnamefont {M.}~\bibnamefont {Prakash}},
  \ and\ \bibinfo {author} {\bibfnamefont {J.~M.}\ \bibnamefont {Lattimer}},\
  }\href {\doibase 10.1103/PhysRevD.82.024016} {\bibfield  {journal} {\bibinfo
  {journal} {Phys. Rev. D}\ }\textbf {\bibinfo {volume} {82}},\ \bibinfo
  {pages} {024016} (\bibinfo {year} {2010})},\ \Eprint
  {http://arxiv.org/abs/1004.5098} {arXiv:1004.5098} \BibitemShut {NoStop}%
\bibitem [{\citenamefont {Ajith}\ and\ \citenamefont
  {Bose}(2009)}]{Ajith:2009fz}%
  \BibitemOpen
  \bibfield  {author} {\bibinfo {author} {\bibfnamefont {P.}~\bibnamefont
  {Ajith}}\ and\ \bibinfo {author} {\bibfnamefont {S.}~\bibnamefont {Bose}},\
  }\href {\doibase 10.1103/PhysRevD.79.084032} {\bibfield  {journal} {\bibinfo
  {journal} {Phys. Rev.}\ }\textbf {\bibinfo {volume} {D79}},\ \bibinfo {pages}
  {084032} (\bibinfo {year} {2009})},\ \Eprint {http://arxiv.org/abs/0901.4936}
  {arXiv:0901.4936 [gr-qc]} \BibitemShut {NoStop}%
\bibitem [{\citenamefont {Helstrom}(1995)}]{Helstrom}%
  \BibitemOpen
  \bibfield  {author} {\bibinfo {author} {\bibfnamefont {C.~W.}\ \bibnamefont
  {Helstrom}},\ }\href@noop {} {\emph {\bibinfo {title} {Elements of signal
  detection and estimation}}}\ (\bibinfo  {publisher} {Prentice-Hall, Inc.},\
  \bibinfo {address} {Upper Saddle River, NJ, USA},\ \bibinfo {year}
  {1995})\BibitemShut {NoStop}%
\bibitem [{aLI(2010)}]{aLIGO_ZDHP}%
  \BibitemOpen
  \href@noop {} {\emph {\bibinfo {title} {{Advanced LIGO anticipated
  sensitivity curves}}}},\ \bibinfo {type} {Tech. Rep.}\ \bibinfo {number}
  {LIGO-T0900288-v3}\ (\bibinfo  {institution} {LIGO Scientific
  Collaboration},\ \bibinfo {address}
  {https://dcc.ligo.org/LIGO-T0900288/public},\ \bibinfo {year}
  {2010})\BibitemShut {NoStop}%
\bibitem [{\citenamefont {Vallisneri}(2008)}]{Vallisneri:2007ev}%
  \BibitemOpen
  \bibfield  {author} {\bibinfo {author} {\bibfnamefont {M.}~\bibnamefont
  {Vallisneri}},\ }\href {\doibase 10.1103/PhysRevD.77.042001} {\bibfield
  {journal} {\bibinfo  {journal} {Phys. Rev. D}\ }\textbf {\bibinfo {volume}
  {77}},\ \bibinfo {pages} {042001} (\bibinfo {year} {2008})},\ \Eprint
  {http://arxiv.org/abs/gr-qc/0703086} {arXiv:gr-qc/0703086} \BibitemShut
  {NoStop}%
\bibitem [{\citenamefont {Rodriguez}\ \emph {et~al.}(2013)\citenamefont
  {Rodriguez}, \citenamefont {Farr}, \citenamefont {Farr},\ and\ \citenamefont
  {Mandel}}]{Rodriguez:2013mla}%
  \BibitemOpen
  \bibfield  {author} {\bibinfo {author} {\bibfnamefont {C.~L.}\ \bibnamefont
  {Rodriguez}}, \bibinfo {author} {\bibfnamefont {B.}~\bibnamefont {Farr}},
  \bibinfo {author} {\bibfnamefont {W.~M.}\ \bibnamefont {Farr}}, \ and\
  \bibinfo {author} {\bibfnamefont {I.}~\bibnamefont {Mandel}},\ }\href
  {\doibase 10.1103/PhysRevD.88.084013} {\bibfield  {journal} {\bibinfo
  {journal} {Phys. Rev.}\ }\textbf {\bibinfo {volume} {D88}},\ \bibinfo {pages}
  {084013} (\bibinfo {year} {2013})},\ \Eprint {http://arxiv.org/abs/1308.1397}
  {arXiv:1308.1397 [astro-ph.IM]} \BibitemShut {NoStop}%
\bibitem [{\citenamefont {Rodriguez}\ \emph {et~al.}(2014)\citenamefont
  {Rodriguez}, \citenamefont {Farr}, \citenamefont {Raymond}, \citenamefont
  {Farr}, \citenamefont {Littenberg}, \citenamefont {Fazi},\ and\ \citenamefont
  {Kalogera}}]{Rodriguez:2013oaa}%
  \BibitemOpen
  \bibfield  {author} {\bibinfo {author} {\bibfnamefont {C.~L.}\ \bibnamefont
  {Rodriguez}}, \bibinfo {author} {\bibfnamefont {B.}~\bibnamefont {Farr}},
  \bibinfo {author} {\bibfnamefont {V.}~\bibnamefont {Raymond}}, \bibinfo
  {author} {\bibfnamefont {W.~M.}\ \bibnamefont {Farr}}, \bibinfo {author}
  {\bibfnamefont {T.~B.}\ \bibnamefont {Littenberg}}, \bibinfo {author}
  {\bibfnamefont {D.}~\bibnamefont {Fazi}}, \ and\ \bibinfo {author}
  {\bibfnamefont {V.}~\bibnamefont {Kalogera}},\ }\href {\doibase
  10.1088/0004-637X/784/2/119} {\bibfield  {journal} {\bibinfo  {journal}
  {Astrophys. J.}\ }\textbf {\bibinfo {volume} {784}},\ \bibinfo {pages} {119}
  (\bibinfo {year} {2014})},\ \Eprint {http://arxiv.org/abs/1309.3273}
  {arXiv:1309.3273 [astro-ph.HE]} \BibitemShut {NoStop}%
\bibitem [{\citenamefont {Damour}\ \emph {et~al.}(2012)\citenamefont {Damour},
  \citenamefont {Nagar},\ and\ \citenamefont {Villain}}]{Damour:2012yf}%
  \BibitemOpen
  \bibfield  {author} {\bibinfo {author} {\bibfnamefont {T.}~\bibnamefont
  {Damour}}, \bibinfo {author} {\bibfnamefont {A.}~\bibnamefont {Nagar}}, \
  and\ \bibinfo {author} {\bibfnamefont {L.}~\bibnamefont {Villain}},\ }\href
  {\doibase 10.1103/PhysRevD.85.123007} {\bibfield  {journal} {\bibinfo
  {journal} {Phys. Rev.}\ }\textbf {\bibinfo {volume} {D85}},\ \bibinfo {pages}
  {123007} (\bibinfo {year} {2012})},\ \Eprint {http://arxiv.org/abs/1203.4352}
  {arXiv:1203.4352 [gr-qc]} \BibitemShut {NoStop}%
\bibitem [{\citenamefont {Lackey}\ \emph {et~al.}(2012)\citenamefont {Lackey},
  \citenamefont {Kyutoku}, \citenamefont {Shibata}, \citenamefont {Brady},\
  and\ \citenamefont {Friedman}}]{Lackey:2011vz}%
  \BibitemOpen
  \bibfield  {author} {\bibinfo {author} {\bibfnamefont {B.~D.}\ \bibnamefont
  {Lackey}}, \bibinfo {author} {\bibfnamefont {K.}~\bibnamefont {Kyutoku}},
  \bibinfo {author} {\bibfnamefont {M.}~\bibnamefont {Shibata}}, \bibinfo
  {author} {\bibfnamefont {P.~R.}\ \bibnamefont {Brady}}, \ and\ \bibinfo
  {author} {\bibfnamefont {J.~L.}\ \bibnamefont {Friedman}},\ }\href {\doibase
  10.1103/PhysRevD.85.044061} {\bibfield  {journal} {\bibinfo  {journal} {Phys.
  Rev.}\ }\textbf {\bibinfo {volume} {D85}},\ \bibinfo {pages} {044061}
  (\bibinfo {year} {2012})},\ \Eprint {http://arxiv.org/abs/1109.3402}
  {arXiv:1109.3402 [astro-ph.HE]} \BibitemShut {NoStop}%
\end{thebibliography}%

\clearpage
\appendix

\makeatletter\onecolumngrid@push\makeatother
\begin{figure*}[htbp]
  \includegraphics[width=\textwidth]{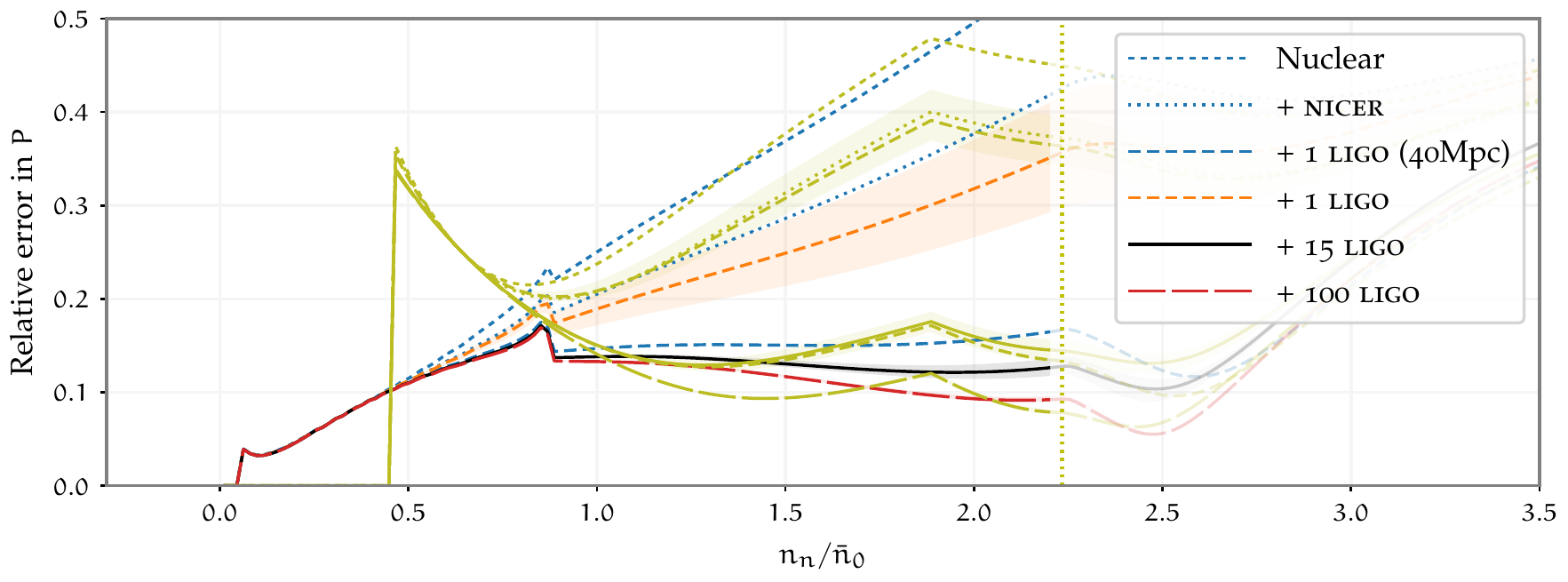}
  \caption{\label{fig:polytrope_comparison}%
    (color online) Relative constraints on the pressure $P$ of nuclear matter in $\beta$-equilibrium for the \gls{Central2c} \gls{EoS} (dark curves) and a polytropic \gls{EoS} (light curves) with the same form as~\cite{Read:2009} with parameters fit to give a similar mass-radius curve, but using the same core \gls{EoS} as ours.
    (Note: Kinks in these curves occur when the form of the \gls{EoS} changes - for example, just below $n=2n_0$, the from of the polytrope~\cite{Read:2009} changes.)
  }
\end{figure*}
\makeatletter\onecolumngrid@pop\makeatother

\section{Supplementary Material}

\subsection{Surface Term in the CLDM}\label{sec:surface-term-cldm}
In our implementation of the \gls{CLDM}, we use the following surface term with
an effective surface tension $\sigma(n_n^i, n_p^i) = \sigma_0/(1 -
  C_{\text{sym}}f(\xp)) \approx \sigma_0(1 - C_{\text{sym}}\beta^2 +
  \order(\beta^4))$ following~\cite{Lattimer:1985} (see also~\cite{Steiner:2012a}):
\begin{subequations}
  \begin{align}
    E & = 4\pi r_p^2 \;
    \sigma_0\frac{1}
    {1 - \frac{C_{\text{sym}}}{96}
      \left(16 - \xp^{-3} - (1-\xp)^{-3}\right)} \\
      & = 4\pi r_p^2 \; \sigma_0\left(
    1 - C_{\text{sym}}\beta^2 + \order(\beta^4)\right),
  \end{align}
  where $\xp = n_p^i/(n_p^i + n_n^i)$ is the proton fraction in the nucleus, and
  $\beta = (n_n^i - n_p^i)/(n_p^i + n_n^i) = 1-2\xp$.  We parameterize
  this as $C_{\text{sym}} = \sigma_\delta/\sigma_0$ where
  $\sigma_\delta \approx \SI{1.38}{MeV/fm^2}$ is held fixed as a
  parameter of the theory, and $\sigma_0$ is varied to smoothly match
  the tabulated outer-crust data.
\end{subequations}

\subsection{Polytropes}
In \cref{fig:polytrope_comparison} we compare the constraints obtained on the total pressure $P(n_B)$ of nuclear matter in $\beta$-equilibrium using our \gls{Central2c} unified parameterization with those obtained using the piecewise polytropic \gls{EoS} in~\cite{Read:2009}.
To better compare these, we do the following:
\begin{enumerate}
  \item Fit the parameters of the polytrope to best match our \gls{Central2c} \gls{EoS}: $\log(p_1) = \num{34.3}$, $\Gamma_1=\num{2.60}$, $\Gamma_2=\num{3.81}$, and $\Gamma_3=\num{2.91}$.
  \item We use the same speed-of-sound core parameterization with $E_c=\SI{350.0}{MeV/fm^3}$, $E_{\mathrm{max}}=\SI{800.0}{MeV/fm^3}$, $C_{\mathrm{max}}=\num{0.8}$ as our \gls{Central2c} \gls{EoS}.
  \item We start with a bare ``Nuclear'' constraint by computing the $1.2\sigma$ 
        covariance matrix of the parameters from Table III of~\cite{Read:2009} over the following \gls{EoS} models that have a small pressure $P(n_0) < \SI{3}{MeV/fm^3}$ at saturation density: \mysc{pal6}, \mysc{sl}y, \mysc{apr1}, \mysc{apr2}, \mysc{apr3}, \mysc{apr4}, \mysc{fps}, \mysc{wff1}, \mysc{wff2}, \mysc{wff3}, \mysc{bbb2}, \mysc{bpal12}, \mysc{eng}, \mysc{mpa1}, \mysc{bgn1h1}, \mysc{pcl2}, \mysc{alf1}, \mysc{alf2}, \mysc{alf3}, and \mysc{alf4}.
        (This excludes some models with hyperon (\mysc{gnh3}, \mysc{h1-7}), pion (\mysc{ps}), and kaon (\mysc{gs1-2}) condensates, as well as the strange-quark matter models \mysc{ms1-2}, which all have significantly higher saturation pressures $P(n_0)> \SI{3}{MeV/fm^3}$).
        This gives similar bare ``Nuclear'' errors as our \gls{Central2c} model at and above saturation density.
\end{enumerate}

We note that the constraints on $P$ are very similar to those from our ``Nuclear'' parameter set.
To obtain this, however, it was critical to use correlated errors in the polytrope parameters.
To this end, taking a polytropic \gls{EoS} with uncorrelated priors is inadvisable.
Only once correlated priors are used does the polytropic equation of state provide constraints comparable to those that can be obtained from our nuclear parameterization.

\subsection{Tabulated EoS Data}
Here we present somewhat tighter constraints on tabulated \gls{EoS} data, required to ensure convexity, than we have seen presented in the literature.
Suppose we have an interval with tabulated density, pressure, and energy $n_{0,1}$, $P_{0,1}$, and $\E_{0,1}$.
If these data come from an equation of state that satisfies thermodynamic convexity $\E''(n) = P'(n)/n \geq 0$ and causality $\E'(P) \geq 1$, then each interval must satisfy the following conditions:
\begin{widetext}
\begin{align}
  P_1 - P_0 &\leq \E_1 - \E_0,
  &\frac{\E_1 + P_1}{\sqrt{(\E_0 + P_0)(E_0 + 2P_1 - P_0)}}
  &\leq \frac{n_1}{n_0} \leq
    \frac{\sqrt{(\E_1 + P_1)(\E_1 + 2P_0 - P_1)}}{\E_0 + P_0}.
\end{align}
\end{widetext}
The tabulated date in~\cite{Sharma:2015} used for the outer crust required some minor corrections to ensure these constraints are met.

\subsection{Thermodynamic Relationships}
Here we briefly review some $T=0$ thermodynamic relationships for an \gls{EoS} with a single conserved component with density $n$ and chemical potential $\mu$, energy density $\mathcal{E}$, energy per particle $E$, and pressure $P$.
These are used at various places throughout the text, such as relating the slope of the symmetry energy $L=3P_n(n_0)/n_0$ to the pressure of neutron matter in \cref{eq:L}
\begin{subequations}
  \begin{gather}
    E(n) = \frac{\mathcal{E}(n)}{n}, \qquad
    \mu = \mathcal{E}'(n),\\
    P = \mu n - \mathcal{E} = n^2E'(n),\\
    C = \frac{c_s^2}{c^2}
    = \diff{P}{\mathcal{E}}
    = \frac{P'(n)}{\mathcal{E}'(n)} = \frac{n\mu'(n)}{\mu}.
  \end{gather}
\end{subequations}

\subsection{Comparison Plots}
On the following pages, we provide comparison plots for all of the \gls{EoS} models discussed in the text.

\begin{figure*}[p]
  \includegraphics[width=\textwidth]{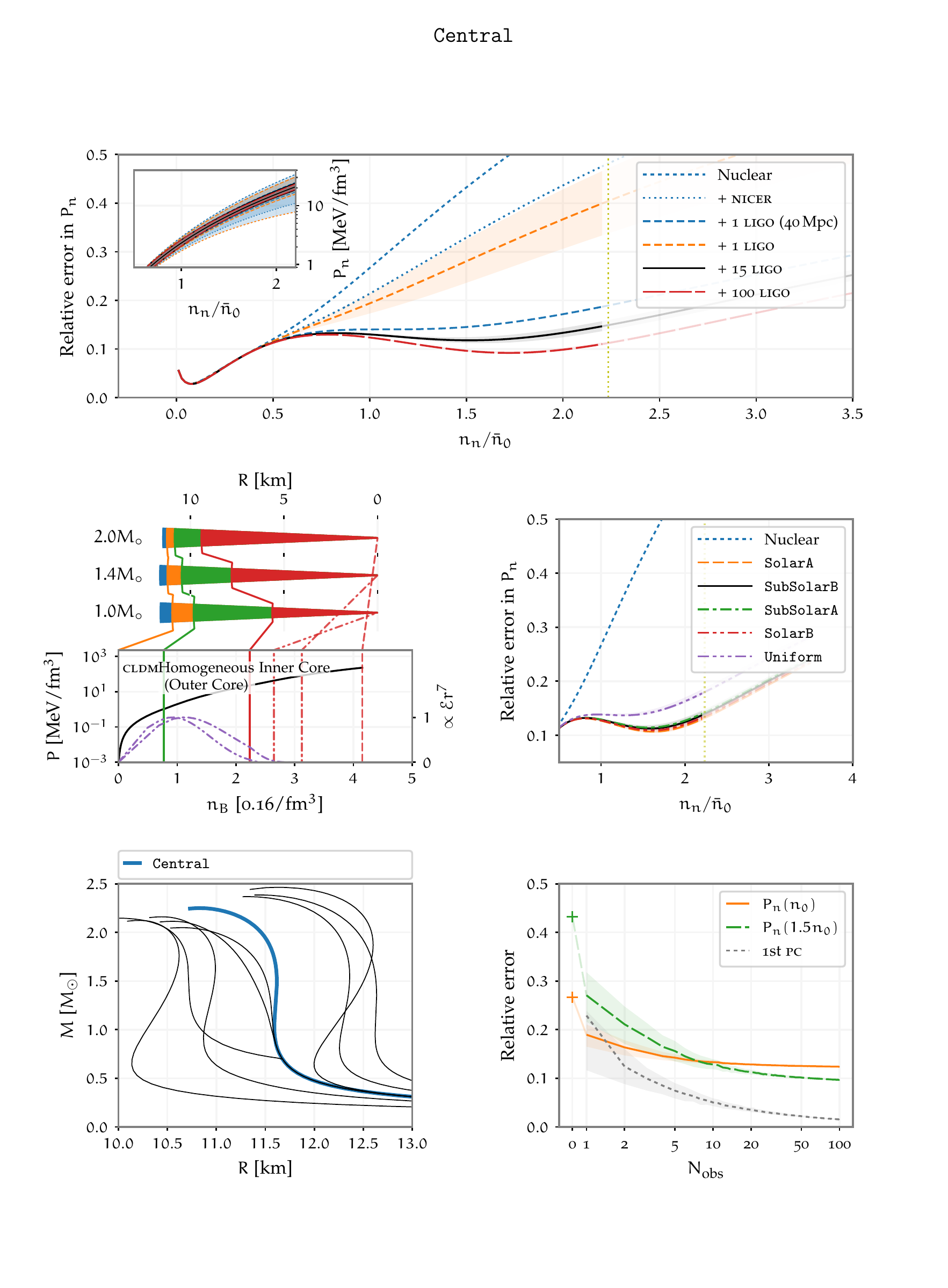}
\end{figure*}
\begin{figure*}[p]
  \includegraphics[width=\textwidth]{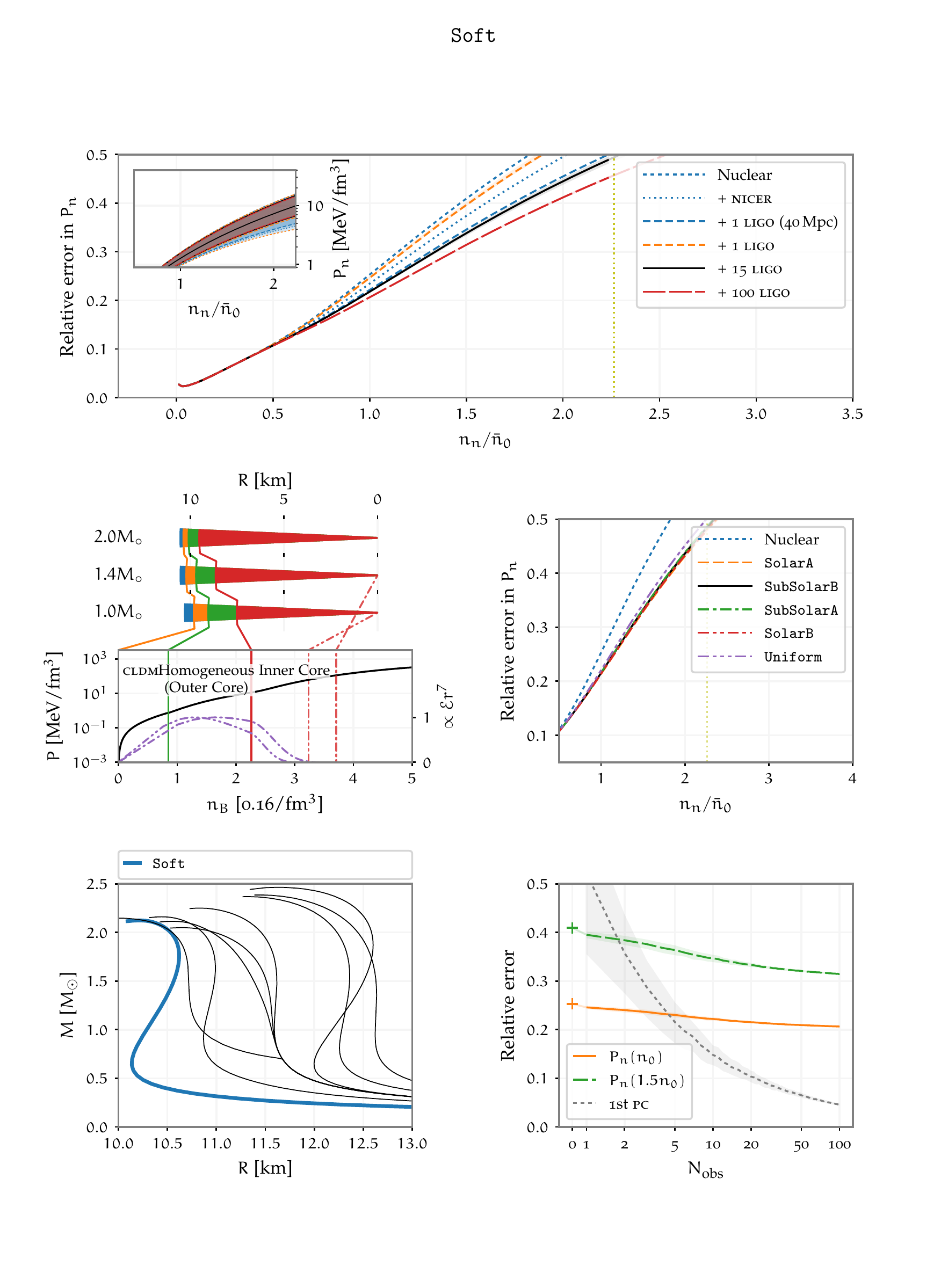}
\end{figure*}
\begin{figure*}[p]
  \includegraphics[width=\textwidth]{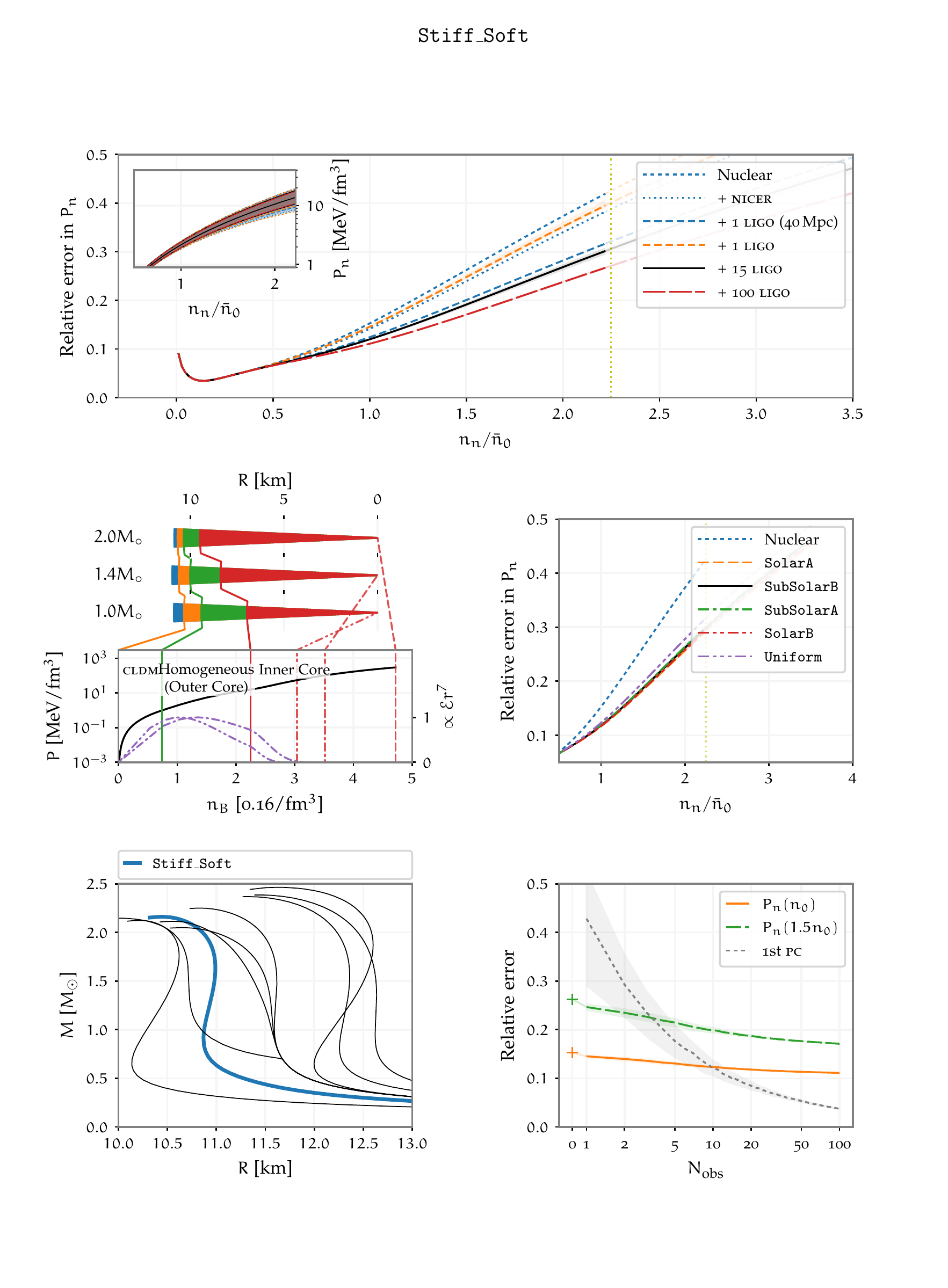}
\end{figure*}
\begin{figure*}[p]
  \includegraphics[width=\textwidth]{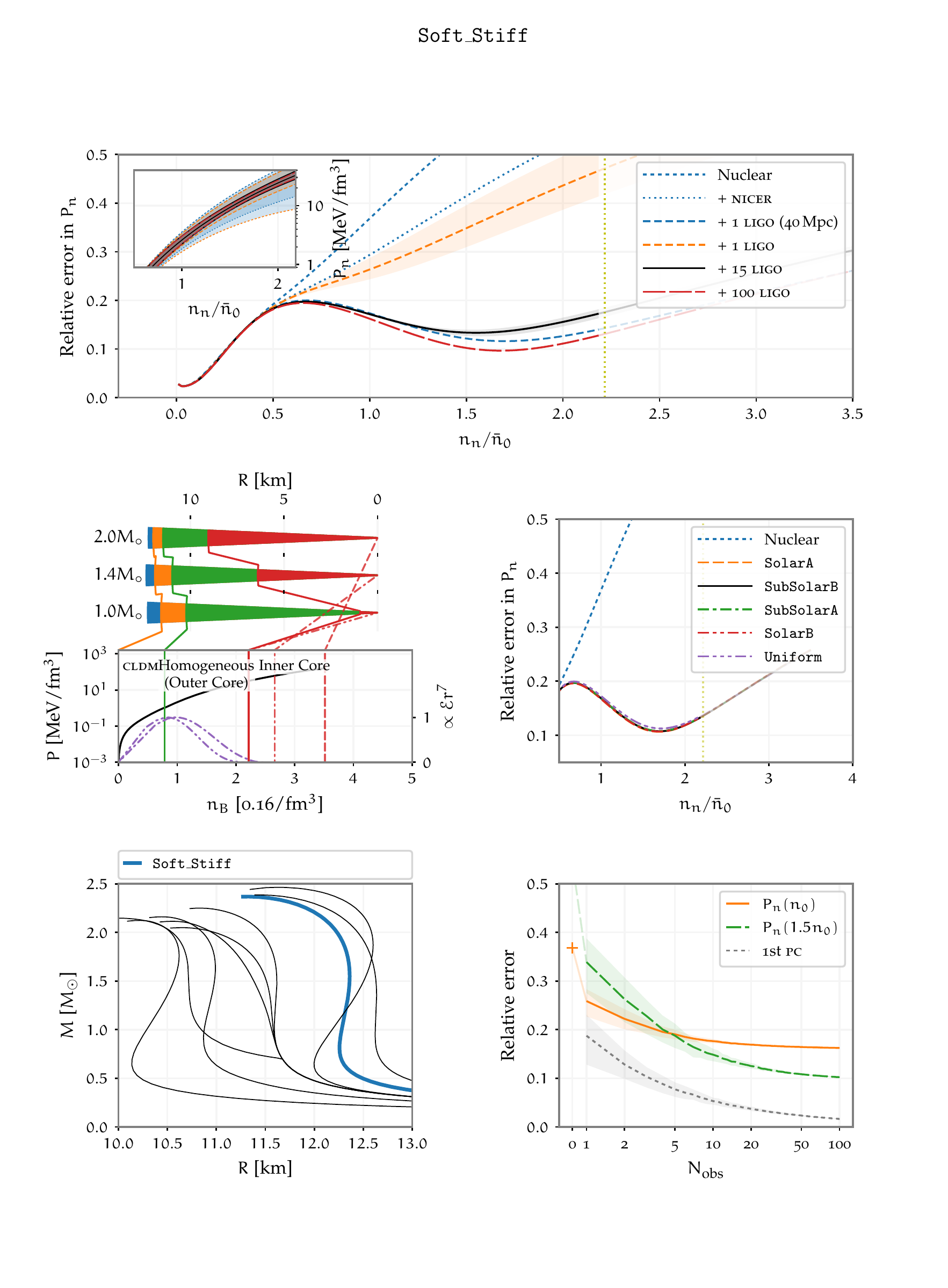}
\end{figure*}
\begin{figure*}[p]
  \includegraphics[width=\textwidth]{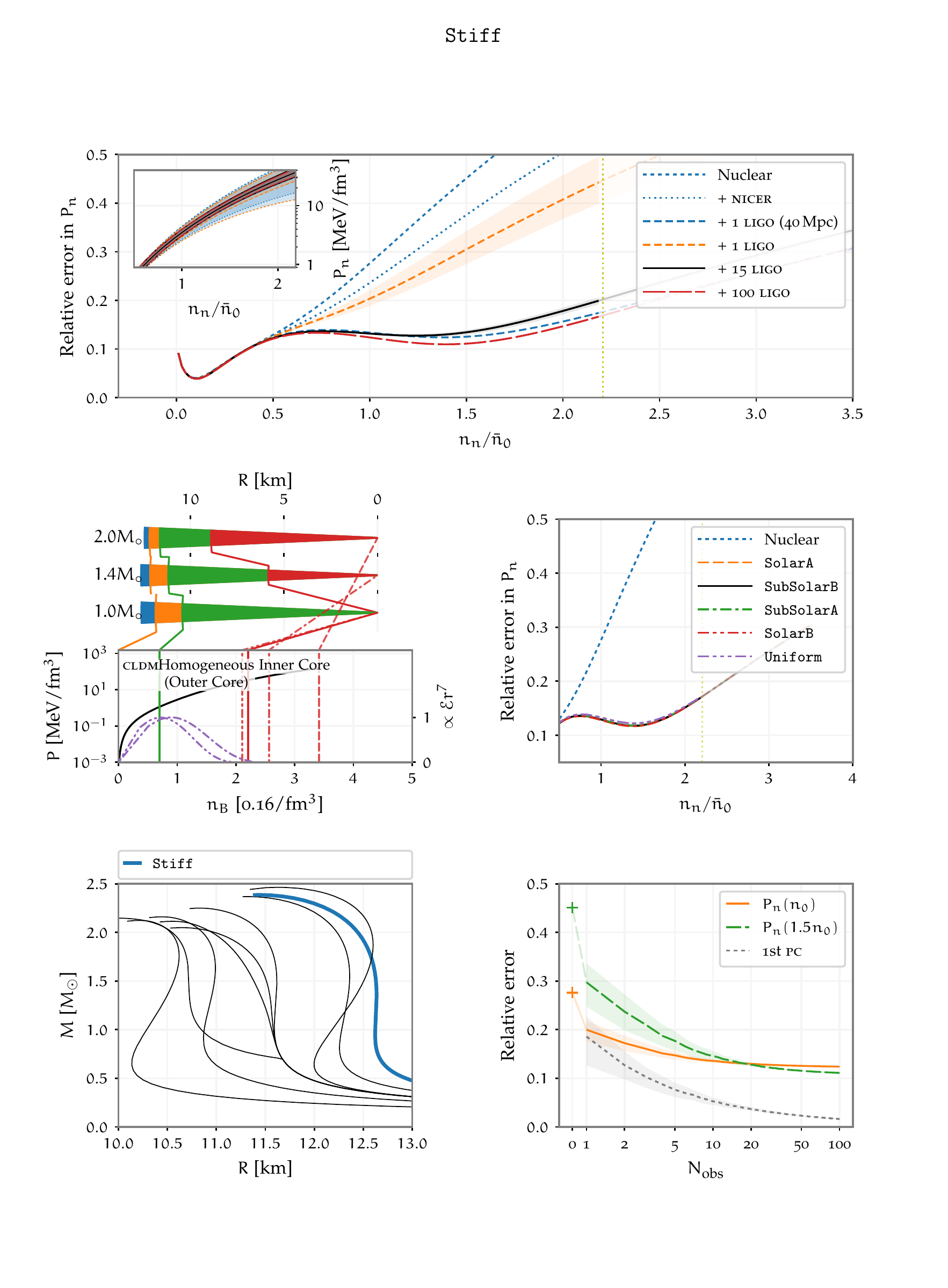}
\end{figure*}
\begin{figure*}[p]
  \includegraphics[width=\textwidth]{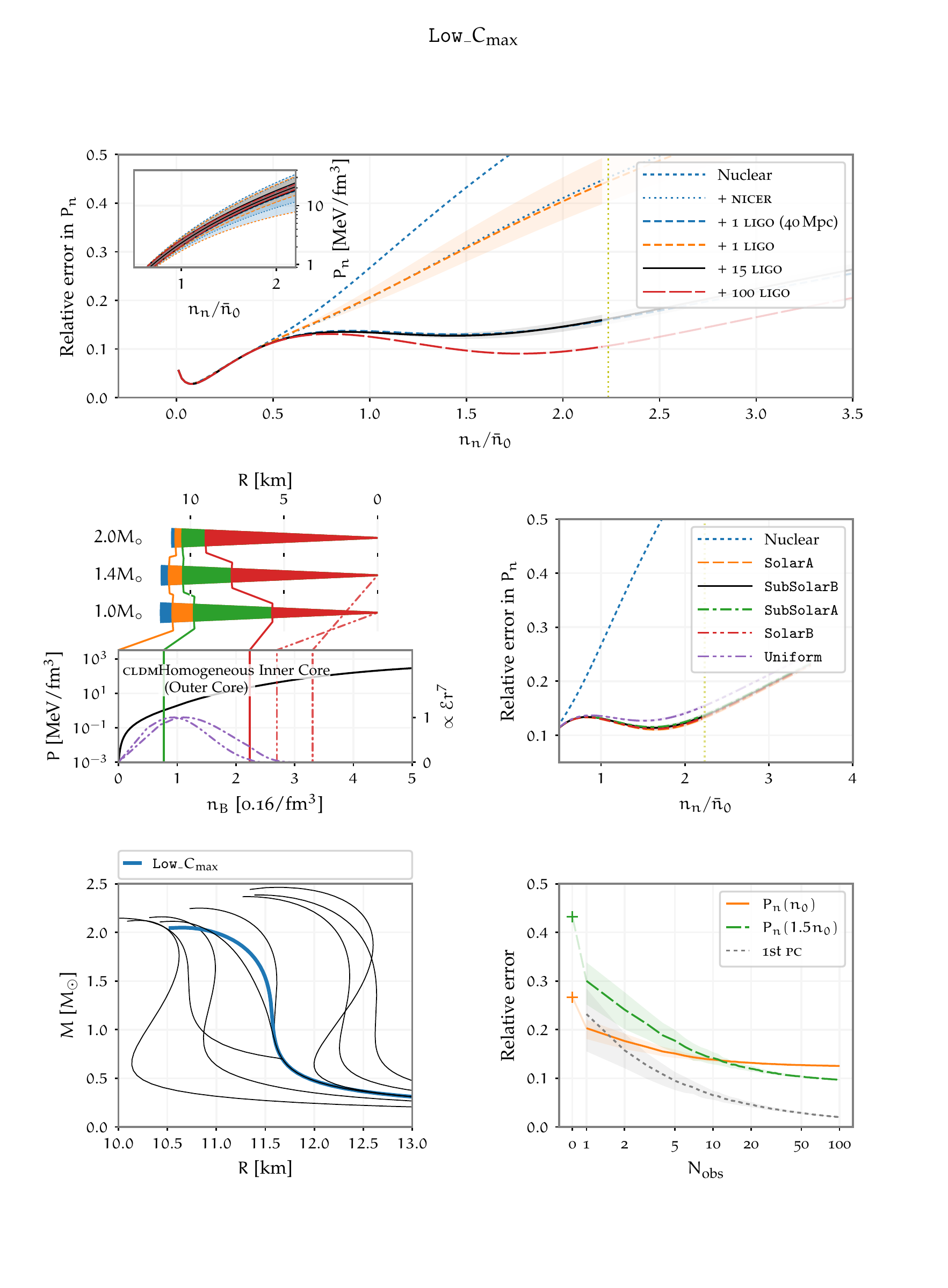}
\end{figure*}
\begin{figure*}[p]
  \includegraphics[width=\textwidth]{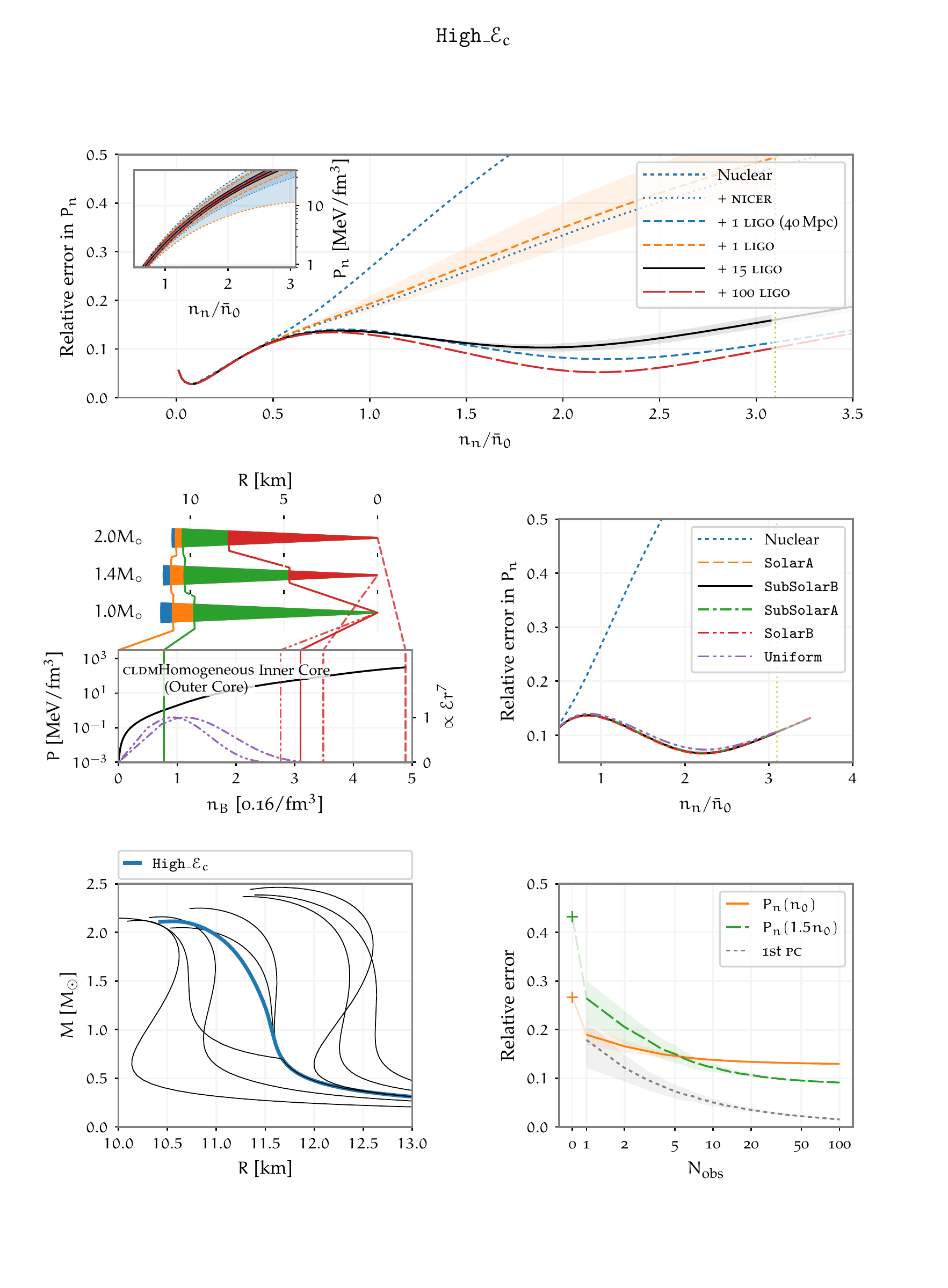}
\end{figure*}
\begin{figure*}[p]
  \includegraphics[width=\textwidth]{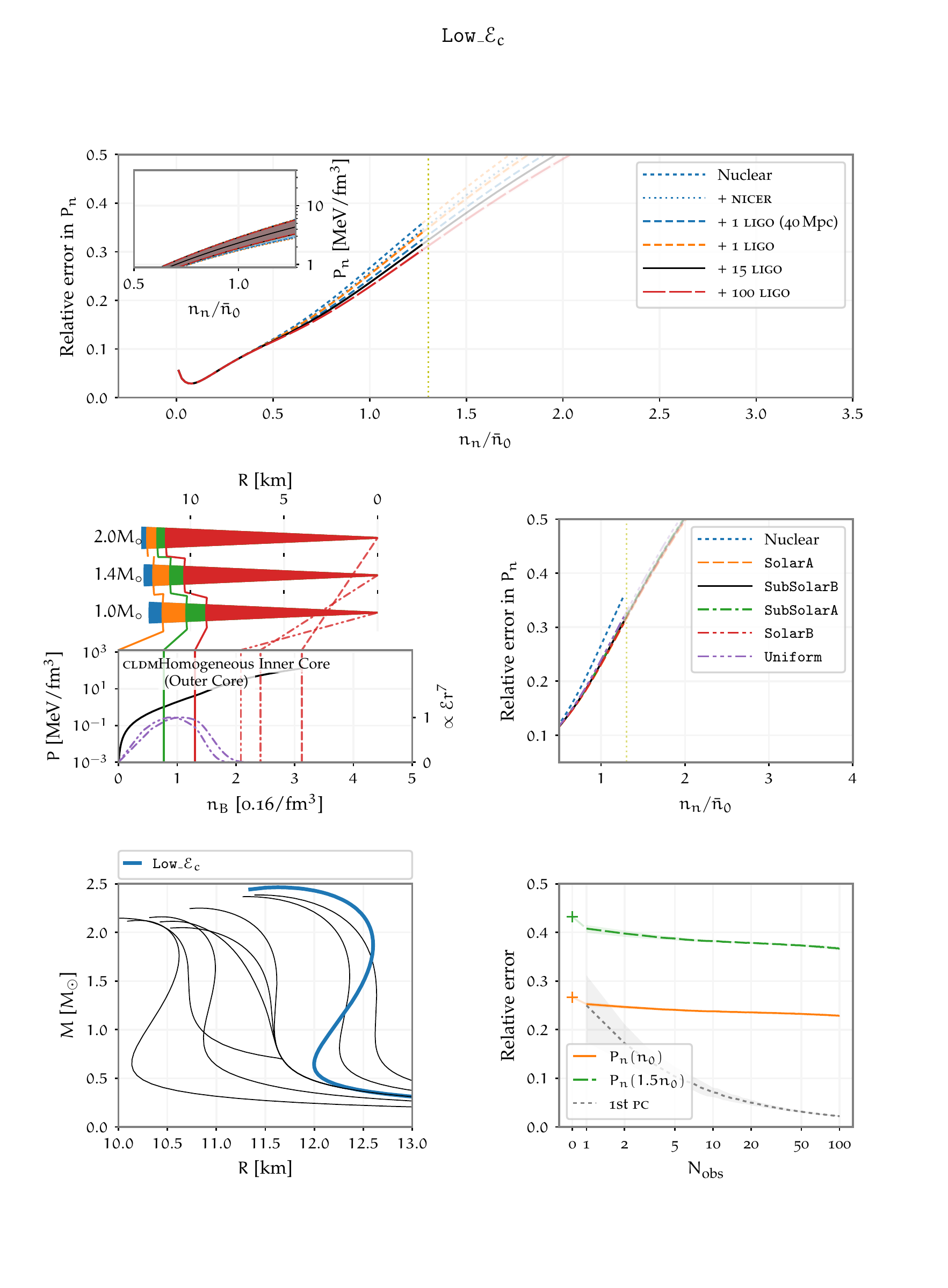}
\end{figure*}
\begin{figure*}[p]
  \includegraphics[width=\textwidth]{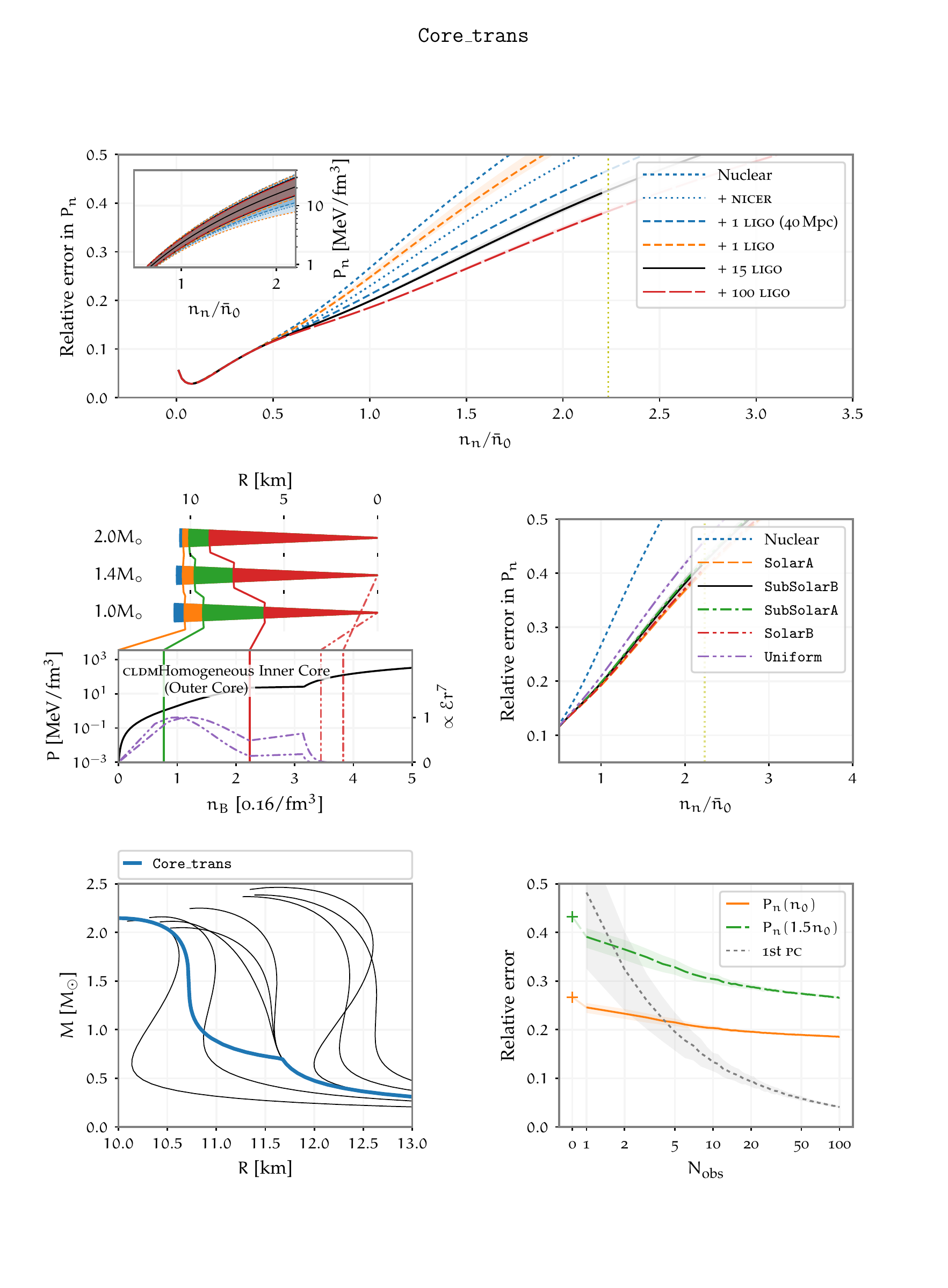}
\end{figure*}

\end{document}